\documentclass[12pt]{article}
\pdfoutput =1

\usepackage{float} 
\usepackage{amsmath}
\usepackage{slashed}
\usepackage{amsfonts}   
\usepackage{amssymb}
\def\a{\alpha}

\def\l{\lambda}
\def\b{\beta}
\def\g{\gamma}

\def\s{\sigma}

\def\no{\nonumber}

\def\diff{d}

\usepackage[a4paper,bindingoffset=0.2in,%
            left=1.8in,right=0.5in,top=1.8in,bottom=0.8in,%
            footskip=.55in]{geometry}

\begin{document}
\date{}
%%%%%%%%%%%%%%%%%%%%

\title{\vspace{-3cm} \hspace{10cm} \small YITP-17-124\\ \vspace{1cm}
{\bf{\Large Fast spinning strings on $ \eta $ deformed $ AdS_5 \times S^{5} $}}}

\author{
{\bf {\normalsize Aritra Banerjee,$^{a}$ Arpan Bhattacharyya,$^{b,c}$ Dibakar Roychowdhury$^{d}$}
\thanks{E-mail: aritra@itp.ac.cn, bhattacharyya.arpan@yahoo.com, dibakarphys@gmail.com}
}
\\
{\normalsize $^{a}$CAS Key Laboratory of Theoretical Physics,}\\{\normalsize Institute of Theoretical Physics,}\\{\normalsize Chinese Academy of Sciences,  Beijing 100190, P.R.China}\\
 {\normalsize $^{b}$ Center for Gravitational Physics, Yukawa Institute for Theoretical Physics,}\\{\normalsize
Kyoto University, Kyoto 606-8502, Japan.}\\{\normalsize $^{c}$ Department of Physics and Center for Field Theory and Particle Physics,}\\{\normalsize Fudan University, 220 Handan Road, 200433 Shanghai, P. R. China.}\\
 {\normalsize $^{d}$ Department of Physics, Swansea University,}\\
  {\normalsize Singleton Park, Swansea SA2 8PP, United Kingdom}
}
%\date{}

\maketitle
%%%%%%%%%%%%%%%%%%%%%%%%%%%%%%%%%%%%%%%%%%%%%%%%%%%%%%%%%%%%%%%%%%%%%%%%%%%%
\begin{abstract}
In this paper, considering the correspondence between spin chains and string sigma models, we explore the rotating string solutions over $ \eta $ deformed $ AdS_5 \times S^{5} $ in the so-called fast spinning limit. In our analysis, we focus only on the bosonic part of the full superstring action and compute the relevant limits on both $(R \times S^{3})_{\eta} $ and $(R \times S^{5})_{\eta} $ models. The resulting system reveals that in the fast spinning limit, the sigma model on $ \eta $ deformed $S^5$ could be \textit{approximately} thought of as the continuum limit of anisotropic $ SU(3) $ Heisenberg spin chain model. We compute the energy for a certain class of spinning strings in deformed $S^5$ and we show that this energy can be mapped to that of a similar spinning string in the purely imaginary $\beta$ deformed background. 
\end{abstract}
%%%%%%%%%%%%%%%%%%%%%%%%%%%%%%%%%%%%%%%
\section{Overview and Motivation}
The quest for integrable deformations \cite{Beisert1}-\cite{Zoubos:2010kh} associated to $ AdS_5 \times S^{5}$ superstring sigma model has been one of the fascinating areas of modern theoretical investigation during the last one decade \footnote{The interested reader can be redirected to \cite{vanTongeren:2013gva} and references therein for a recent introduction to the subject.}. The fact that  $ AdS_5 \times S^{5}$ can be represented by a supercoset and the string sigma model action constructed in terms of the coset group elements can be shown to be classically integrable \cite{Bena:2003wd}, has spawned a plethora of studies based on deformation of the supercoset. This, in turn, has generated a lot of attention toward geometric deformations of target spaces associated to two-dimensional deformed integrable sigma models.  Recently, the class of Yang-Baxter (YB) deformations \cite{Klimcik:2002zj}-\cite{Kawaguchi:2014fca} of $ AdS_5 \times S^{5} $ superstring sigma model has gained renewed attention due to its several remarkable properties namely, the existence of Lax connection and the fermionic kappa symmetry. This new class of integrable deformation is known as the $ \eta $- deformed sigma model where $ \eta $ is the deformation parameter that enters into the classical currents which still satisfies the (modified) classical Yang Baxter equations \footnote{For recent investigations on $\eta$ deformed models from the point of view of Seiberg -Witten maps please see \cite{gandu, gandu1}.}.

Unlike the existing plane wave limit \cite{Blau:2001ne}-\cite{Plefka:2003nb} for type IIB strings propagating in $ AdS_5 \times S^{5} $, the corresponding pp wave limit for $ \eta $ deformed  $ AdS_5 \times S^{5} $ is still lacking in the literature. As a consequence of this, it is completely unclear at the moment whether one could talk about anything like perturbative plane wave string/gauge theory duality like in the case for $ AdS_5 \times S^{5} $ superstrings \cite{Beisert:2003ea}.

The purpose of the present article is basically to address this issue following a different path and to understand the $ \eta  $ deformed sigma models in the light of spin chain/string sigma model correspondence \cite{Arutyunov:2003uj}-\cite{Kameyama:2013qka}. Following the original construction \cite{Kruczenski:2003gt}, in the present analysis we consider  rotating string configurations on deformed spheres, i.e.  $(R \times S^{3})_{\eta} $ and  $ (R \times S^{5})_{\eta} $ in the fast spinning limit \cite{Hernandez:2004uw} which is analogous to that of taking a BMN limit as considered by authors in \cite{Berenstein:2002jq}.  At this stage it is noteworthy to mention that the anisotropic Landau-Lifshitz equations corresponding the bosonic sector of $ \eta $ deformed superstrings had already been explored by authors in \cite{Kameyama:2014bua} where they had mapped the deformed $SL(2)$ and $ SU(2) $ sector of the spin chain to that with the fast spinning limit associated to string sigma models on time-like warped $ AdS_3\times S^{1} $ and $ R \times $ squashed $ S^{3}. $\footnote{For other discussions on fast spinning limits of $\eta$ deformed models and associated Neumann-Rosochatius systems, one could look at \cite{Arutyunov:2014cda, Arutyunov:2016ysi, Kameyama:2014vma, Hernandez1}. } This limit, it seems, only could be consistently taken when the deformation parameter is small.   In  our analysis after taking the fast spinning limit, the $ \eta $ deformed sigma model corresponding to $ (R \times S^{3})_{\eta} $ could be thought of as being that of the continuum limit of XXZ $ SU(2) $ Heisenberg spin chain (which is a well known integrable model \cite{Frolov:2005iq}). 

For the deformed five-sphere case it is harder to play around with integrability for the larger sector. The usual bosonic symmetry group of $SO(6)$ has been broken down to $U(1)\times U(1)\times U(1)$ isometries for the $\eta$ deformed case. Still, the classical integrability of the full background is supposed to be inherited in the deformed theory too. This background has added nontrivial $NS-NS$ fields associated to the sigma model. A fast spinning string limit on this generates the usual $SU(3)$ spin chain terms with anisotropies, sans terms that cannot be reproduced via $SU(3)$ coherent state components up to phases and modulo the exact two form field of the associated string sigma model. In other words, the full $ \eta $ deformed sigma model, which is expected to have a quantum group ($q$-deformed) symmetry, cannot be expressed in terms of standard $ SU(3) $ coherent vectors. The complete understanding of the $q$-deformed coherent states, that is expected to resolve this conflict, remains to be a puzzle at the moment which we leave for the purpose of future investigations. 

The organization of the paper is the following. In Section 2, we revisit the sigma model corresponding to fast spinning strings on $ (R \times S^{3})_{\eta} $. Keeping the spirit of the earlier analysis \cite{Kruczenski:2003gt}, we express the spin chain Hamiltonian in terms of usual $ SU(2) $ coherent states. It turns out that $ \eta $ deformations add non trivially to the spin chain Hamiltonian in such a way so that it takes the form of an anisotropic Heisenberg spin chain having structural similarities to the XXZ model. Finally, we calculate the energy corresponding to a particular class of stringy configuration namely the circular string solutions associated to anisotropic $ SU(2) $ spin chain model (with periodic boundary conditions) where we find the leading order correction to the energy which turns out to be quadratic in the deformation parameter.

In Section 3, we compute the sigma model corresponding to fast spinning strings on $ (R\times S^{5})_{\eta} $, which also turns out to have a structure similar to the anisotropic $SU(3)$ spin chain with an added correction term quadratic in the deformation parameter.  Then we explicitly show that it is not possible to incorporate a correction term having a product of three components of $SU(3)$ coherent state vector via the usual nearest-neighbor interaction spin chain picture. Also, the only contribution due to the $ B $ field can't be taken care of explicitly using this approach, which is not entirely unexpected since the underlying $ SU(3) $ symmetry is broken in the presence of nontrivial background $NS$ fluxes. From the above discussions, it should also be quite evident that it is the presence of background $B$ field that creates a clear distinction between the two sigma models corresponding to $ (R\times S^{3})_{\eta}  $ and $  (R\times S^{5})_{\eta} $ from the perspective of a spin chain. 

In section 4, we discuss how the integrable models corresponding to fast spinning string in subsectors of $(AdS_5\times S^5)_{\eta}$ can be mapped to other known deformed models in the literature. It is particularly interesting to see that the Yang-Baxter sigma models arose out of studies to generalize sigma models on non-symmetric cosets, for example, that of on the squashed spheres. A closely related example of is that of $\gamma$ and $\beta$ deformed backgrounds (\cite{Lunin:2005jy},\cite{Frolov:2005dj}) constructed via series of T dualities and shifts. We comment on these connections at the level of underlying continuum limits of spin chain picture. We also unearth a surprising similarity between the purely imaginary $\b$ deformed Lunin-Maldacena background and the $\eta$ deformed $SU(3)$ theories in the fast spinning limit in relation to the spinning string solutions in these two. After discussing further open problems, we conclude our analysis in section 5.

%%%%%%%%%%%%%%%%%%%%%%%%%%%%%%%%%%%%%%%%
\section{Revisiting fast spinning strings on $ (R \times S^{3})_{\eta}$ }
The purpose of this section is to revisit the sigma model corresponding to fast spinning strings on $ (R \times S^{3})_{\eta} $ subsector of the total deformed spacetime and to take a worldsheet fast spinning string limit. In the case of the undeformed three-sphere  \cite{Kruczenski:2003gt} this non relativistic limit maps the system of classical strings propagating in the background to the continuum limit of the $ SU(2) $ Heisenberg spin chain Hamiltonian. 
%%%%%%%%%%%%%%%%%%%%%%%%%%%%%%%%%%%%%%%%%%%%%%%%%%%%%
\subsection{The sigma model}
The $ \eta $ deformed $ AdS_3 \times S^{3} $ background could be formally expressed as
%\footnote{We have set, $ L=1 $ in our analysis.} 
\cite{Hoare:2014pna},
\begin{eqnarray}
ds^{2}_{AdS_3 \times S^{3}}= -\mathfrak{h}(\varrho)dt^{2}+\mathfrak{f}(\varrho)d\varrho^{2}+\varrho^{2}d\psi^{2}+ \tilde{\mathfrak{h}}(\theta)d\varphi^{2}+\tilde{\mathfrak{f}}(\theta)d\theta^{2}+\cos^{2}\theta d\phi^{2}
\label{E1}
\end{eqnarray}
where, the functions in the metric components above could be explicitly written as,
\begin{eqnarray}
\mathfrak{h}&=&\frac{1+\varrho^{2}}{(1-\kappa^{2}\varrho^{2})},~~\mathfrak{f}=\frac{1}{(1+\varrho^{2})(1-\kappa^{2}\varrho^{2})}\nonumber\\
\tilde{\mathfrak{h}}&=&\frac{\sin^{2}\theta}{(1+\kappa^{2}\cos^{2}\theta)},~~\tilde{\mathfrak{f}}=\frac{1}{(1+\kappa^{2}\cos^{2}\theta)}.\label{E2}
\end{eqnarray}

Notice that here the parameter $ \kappa $ is related to the original deformation parameter $ \eta $ as \cite{Arutyunov:2013ega},
\begin{eqnarray}
 \kappa = \frac{2\eta}{1-\eta^{2}}.
\end{eqnarray}
Henceforth, we would denote $ \kappa $ as being the deformation parameter in our analysis.

In order to proceed further, we fix the coordinates on deformed $ AdS_3 $ and consider rotating closed strings propagating in $ (R \times S^{3})_{\eta} $ where we choose an ansatz of the following form,
\begin{eqnarray}
t= \xi \tau, ~\varrho =0,~\theta =\theta (\sigma, \tau),~\varphi\ =\varphi (\sigma ,\tau), ~\phi = \phi (\sigma ,\tau).
\end{eqnarray}
where, $ \xi $ is the energy associated with classical stringy configuration and ($ \sigma ,\tau $ ) are the world-sheet coordinates. Following the original prescription \cite{Kruczenski:2003gt}, we consider the change of coordinates on deformed $ S^{3} $ as,
\begin{eqnarray}
\phi =\xi \tau +\Phi_{1}+\Phi_{2} ,~~\varphi = \xi \tau +\Phi_{1}-\Phi_{2}.\label{E5}
\end{eqnarray}

Substituting (\ref{E5}) into (\ref{E1}), the relevant part of the background metric turns out to be,
\begin{eqnarray}
ds^{2}_{R\times S^{3}}=-\xi^{2}d \tau^{2}+\tilde{\mathfrak{h}}(\theta)( \xi d\tau +d\Phi_{1}-d\Phi_{2})^{2}+\tilde{\mathfrak{f}}(\theta) d\theta^{2}+\cos^{2}\theta( \xi d\tau +d\Phi_{1}+d\Phi_{2})^{2}.
\end{eqnarray}

Considering the conformally flat world-sheet metric, the corresponding Polyakov Lagrangian could be formally expressed as\footnote{Notice that, here $ \hat{\lambda}=\lambda (1+\kappa^{2})$ corresponds to the modified string tension \cite{Hoare:2014pna}.},
\begin{eqnarray}
\mathcal{L}_{P}=\frac{\sqrt{\hat\lambda}}{4\pi}\Big[ g_{\tau \tau}+g_{\Phi_{1}\Phi_{1}}(\dot{\Phi}^{2}_{1}-\Phi'^{2}_{1})+g_{\Phi_{2}\Phi_{2}}(\dot{\Phi}^{2}_{2}-\Phi'^{2}_{2})+g_{\theta \theta}(\dot{\theta}^{2}-\theta'^{2})\nonumber\\
+2g_{\tau \Phi_{1}}\dot{\Phi}_{1}+2g_{\tau \Phi_{2}}\dot{\Phi}_{2}+2g_{\Phi_{1}\Phi_{2}}(\dot{\Phi}_{1}\dot{\Phi}_{2}-\Phi'_{1}\Phi'_{2}) \Big]. \label{E7}
\end{eqnarray}

Our next step would be to take the large spin limit \cite{Kruczenski:2003gt}-\cite{Kameyama:2013qka} corresponding to the rotating string configuration where we set, $ \xi \rightarrow \infty$ such that both $ \xi \dot{X}^{\mu} $ as well as $ \xi^{2}\kappa^{2} $ is finite. In other words, the fast spinning limit corresponds to setting both, $ \dot{X}^{\mu}, \kappa^{2}\rightarrow 0 $. One should note here that the fast spinning string limit additionally constrains the values of the deformation parameter, which enables us to work in the leading order of $\kappa$ throughout the calculation.

Finally, considering the large spin limit the corresponding sigma model Lagrangian (\ref{E7}) turns out to be (sans a total derivative term),
\begin{eqnarray}
\mathcal{L}_{P}(1+\kappa^2\cos^2\theta)&=&\frac{\sqrt{\hat\lambda}}{4\pi}\Big[-\xi^{2}\kappa^{2}\sin^{2}\theta \cos^{2}\theta +(1+\kappa^2\cos^4\theta)(-\Phi'^{2}_{1}-\Phi'^{2}_{2}+2\xi \dot{\Phi}_{1}) \nonumber\\
&-&\theta'^{2}+(\kappa^2\cos^4\theta+\cos 2\theta)(2\xi \dot{\Phi}_{2}  -2 \Phi'_{1}\Phi'_{2})\Big]\label{E9}
\end{eqnarray}
which could be further truncated in the small deformation regime as,
\begin{eqnarray}
\mathcal{L}_{P}=\frac{\sqrt{\lambda}}{4\pi}\Big[-\xi^{2}\kappa^{2}\sin^{2}\theta \cos^{2}\theta -\Phi'^{2}_{1}-\Phi'^{2}_{2}+2\xi \dot{\Phi}_{1}-\theta'^{2}
+\cos 2\theta(2\xi \dot{\Phi}_{2}  -2 \Phi'_{1}\Phi'_{2})\Big]\label{E9}
\end{eqnarray}
where, one should notice that in the large $\lambda$ limit, the term $\sqrt{\lambda}\xi^2\kappa^2$ still remains finite. 
Next, we compute the angular momentum and identify the leading order term,
\begin{eqnarray}
J_{\Phi_{1}}\sim \frac{2\xi R^{2}(1+\kappa^{2})^{1/2}}{4 \pi \alpha'}\int_{0}^{2 \pi}d \sigma =\frac{\xi R^{2}(1+\kappa^{2})^{1/2}}{\alpha'}=\sqrt{\lambda}\xi
\end{eqnarray}
which clearly diverges in the limit, $ \xi \rightarrow \infty $ and thereby makes sense of what we call as fast spinning limit. Notice that, here we have dropped the subleading term which is of the order $\sim \sqrt{\lambda}\xi \kappa^2$. 

Next, we use one of the Virasoro constraints,
\begin{eqnarray}
T_{\sigma \tau}=0
\end{eqnarray}
in order to eliminate $ \Phi'_{1} $ which yields in the leading order of $\kappa$,
\begin{eqnarray}
\Phi'_{1}=-\cos 2\theta \Phi'_{2}.\label{E11}
\end{eqnarray}

Substituting (\ref{E11}) into (\ref{E9}) we finally write down the Polyakov action in the limit,
\begin{eqnarray}
\mathcal{S}_{P}&=&\frac{\sqrt{\lambda}}{4\pi} \int d\tau d\sigma \left(2\xi \dot{\Phi}_{1}+2\xi \dot{\Phi}_{2} \cos 2\theta -\theta'^{2}-\Phi'^{2}_{2}\sin^{2}2\theta - \xi^{2}\kappa^{2}\sin^{2}\theta \cos^{2}\theta\right) \nonumber\\
&=&\int dt d\tilde{\sigma} \left(\partial_{t} \Phi_{2} \cos 2\theta -\frac{\lambda}{8 \pi^{2}}(\theta'^{2}+\Phi'^{2}_{2}\sin^{2}2\theta) - \frac{\kappa^{2}}{2}\sin^{2}\theta \cos^{2}\theta\right)
\end{eqnarray}
where, in the second line we have ignored the total derivative term (i.e. integral over $\xi \dot{\Phi}_{1}$) and rescaled the coordinate $ \sigma $ as \cite{Kruczenski:2003gt},
\begin{eqnarray}
\tilde{\sigma}=\frac{\sqrt{{\lambda}}\sigma\xi}{2 \pi}=\frac{J \sigma}{2 \pi}\label{E13}
\end{eqnarray}
where we define the angular momentum as $ J_{\Phi_1}=J $.

We can introduce the new variables,
\begin{eqnarray}
\Phi_{2}=-\frac{\Phi}{2},~~\theta =\frac{\Theta}{2}
\end{eqnarray}
and the sigma model Lagrangian could be formally expressed as,
\begin{eqnarray}
\mathcal{S}_{P}=-\int dt d \tilde{\sigma}\left( \frac{1}{2}\cos \Theta \partial_{t}\Phi +\frac{{\lambda}}{32 \pi^{2}}(\Theta'^{2}+\Phi'^{2}\sin^{2}\Theta) + \frac{\kappa^{2}}{8}\sin^{2}\Theta  \right).\label{E15}
\end{eqnarray}
Using (\ref{E13}), one could re-express the sigma model as,
\begin{eqnarray}\label{actionsu2}
\mathcal{S}_{P}=-\frac{J}{2 \pi}\int dt d \sigma\left( \frac{1}{2}\cos \Theta \partial_{t}\Phi +\frac{{\lambda}}{8 J^{2}}(\Theta'^{2}+\Phi'^{2}\sin^{2}\Theta) + \frac{\kappa^{2}}{8}\sin^{2}\Theta  \right)\label{e17}
\end{eqnarray}
where, the ratio $ \frac{{\lambda}}{J^{2}} $ is held fixed in the limit, $ J \rightarrow \infty $. Clearly in the large spin limit the path integral is dominated by its classical saddle point.

A few important points are to be noted at this stage. First of all, in the limit of the vanishing deformation, the above sigma model (\ref{E15}) precisely corresponds to the continuum limit of long Heisenberg spin chains \cite{Kruczenski:2003gt}. However, in the presence of $ \eta $- deformations this spin chain gets deformed in a non trivial fashion where the contribution due to background deformations could be encoded in the deformed Hamiltonian,
\begin{eqnarray} \label{defspin}
\mathcal{H}_{D}=\frac{\kappa^{2}}{8}\int_{0}^{J} \sin^{2}\Theta ~ d \tilde{\sigma}.\label{E17}
\end{eqnarray}
 Before we end this discussion, it is indeed noteworthy to mention that, in (\ref{E2}) the polar angles $\phi$ and $\varphi$ are not equivalent coordinates as is the case for an undeformed three-sphere. In fact one can show that the two 2-spheres on $\phi$ and $\varphi$ are related to each other via a discrete $\mathbb{Z}_2$ symmetry in this case \cite{Hoare:2014pna}.  We can choose the ansatz in the opposite way,
\begin{eqnarray}
\varphi =\xi \tau +\Phi_{1}+\Phi_{2} ,~~\phi = \xi \tau +\Phi_{1}-\Phi_{2}\label{E19}
\end{eqnarray}
which gives rise to the  following string Lagrangian after taking the relevant BMN limit 
\begin{eqnarray} \label{refgnd}
\mathcal{L}_{P}=\xi^{2}\kappa^{2}\sin^{2}\theta \cos^{2}\theta -\Phi'^{2}_{1}-\Phi'^{2}_{2}+2\xi \dot{\Phi}_{1}-\theta'^{2}
-\cos 2\theta(2\xi \dot{\Phi}_{2}  -2 \Phi'_{1}\Phi'_{2}).\label{E20}
\end{eqnarray}

Notice that (\ref{E20}) differs with (\ref{E9}) only by a sign in front of $\cos 2\theta$. Now we can again use the Virasoro constraint as before to find the condition in the leading order of $\kappa$,
\begin{eqnarray}
\Phi'_{1}=\cos 2\theta \Phi'_{2},\label{E21}
\end{eqnarray}
and substitute it back into (\ref{E20}) which finally yields the Lagrangian (\ref{E13}). This is a nice fact to note that in the fast spinning string limit, the asymmetry between the $\phi$ and $\varphi$ spheres goes away. Since in  \cite{Kameyama:2014bua} the fast spinning string in the deformed three-sphere was shown to be agreeing with the Landau Lifshitz sigma model on a squashed sphere \cite{Wen:2006fw}, we can say that the fast-moving string does not discriminate between two different modes of squashing. We would elaborate on this point in a later section.
%%%%%%%%%%%%%%%%%%%%%%%%%%%%%%%%%%%%%%%%%%%%%%%%
\subsection{The anisotropic $ SU(2) $ spin chain}
We now turn to establishing a precise connection between the string sigma model (\ref{E17}) and that of the continuum limit of $ SU(2) $ spin chain Hamiltonian with some nontrivial corrections to it. In our analysis, we follow the same spirit as that of the original analysis \cite{Kruczenski:2003gt}.  One can always rewrite the contribution (\ref{E17}) as ($\kappa^2 - \kappa^2 \cos^{2}\Theta $) and discard the constant term without the loss of generality.  

In order to proceed further, we first define $SU(2)$ coherent state in the spherical polar coordinates. To do so, let us consider the following unit vector,
\begin{eqnarray}
\vec n=\Big\{\sin \Theta\,\sin \Phi ,\sin \Theta\,\cos \Phi ,\cos \Theta \Big\}.
\end{eqnarray}
Next we note down the coherent state corresponding to the spin at the $k^{th}$ site \cite{Perelomov:1986tf},
\begin{equation}
|\vec n_{k}\rangle=D(\vec n_{k})|0\rangle
\end{equation}
where, we define the matrix,
\begin{equation}
D(\vec n_{k})=e^{\bf{{\alpha_{k,\,+}S_{k,\,+} - \,\alpha_{k,\,-}S_{k,\,-}}}},~~\mathbf{\alpha_{k, \pm}}=\frac{\Theta_{k}}{2} e^{\pm i\,\Phi_{k}}
\end{equation}
together with the spin operator defined as,  
\begin{equation}
S_{k,z}|0\rangle=\frac{1}{2}|0\rangle.
\end{equation}
These spin operators could be defined in terms of Pauli matrices,
\begin{equation}
S_{x}=\frac{\sigma_x}{2},\,\,S_{y}=\frac{\sigma_y}{2},\,\,S_{z}=\frac{\sigma_z}{2},\,\,S_{+}=S_{x}-i\,S_{y},\,\,S_{-}=S_{x}+i\,S_{y}.
\end{equation}
Finally, the  coherent state corresponding to the full spin chain could be formally expressed as,
\begin{equation}
|\vec n\rangle=\prod_{k=1}^{L}|\vec n_{k}\rangle,
\end{equation}
where, $L$ stands for the total number of sites on the chain. 

Our next task would be to switch to  the discrete picture. In the continuum limit which corresponds to setting the lattice spacing $a=\frac{1}{L}\sim 0$ we get,
\begin{eqnarray}
\textbf{n}_{k}\rightarrow\textbf{n} (\sigma)=\textbf{n}\left(\frac{k}{L} \right) , \sum_{k=1}^{L}\rightarrow L\,\int_{0}^{2\pi}d\sigma.\label{E28}
\end{eqnarray}\par
Based on (\ref{E28}), we propose the following discrete version corresponding to  (\ref{defspin}) 

\begin{equation}
\mathcal{H}_{D}=\frac{J\,\kappa^2}{4\pi }\sum_{k=1}^{L}\langle\vec n|S_{k,z}S_{k+1,z}|\vec n\rangle.
\end{equation}
where we had used the fact \footnote{  As the term in  (\ref{defspin}) has no derivative, one might be tempted to consider it as a ultra-local term, for example product of two spin operators at a single site, i.e say $S_{k,z}^2.$  As we have used Pauli matrices for various $S_k$, this possibility will be automatically ruled out as the product of two Pauli matrices acting at the same site  will give either identity or a single Pauli matrix.} ,
\begin{equation}
\langle\vec n_{k}| S_{k,z}|\vec n_{k}\rangle\langle\vec n_{k+1}| S_{k+1,z}|\vec n_{k+1}\rangle= \cos \Theta_{k} \cos \Theta_{k+1}.
\end{equation}
%Then we taylor expand $\theta_{k+1}$ around $\theta_{k}$ and keep the leading order term in $L\rightarrow \infty.$
The total Hamiltonian could be formally expressed as,
\begin{eqnarray}
\mathcal{H}_{P}=\frac{J}{2 \pi}\sum_{k=1}^{L}\langle\vec n|\left[\frac{{\lambda}}{J^{2}}\Big(\frac{1}{4}-S_{k,\,x}S_{k+1,\,x}-S_{k,\,y}S_{k+1,\,y}-S_{k,\,z}S_{k+1,\,z}\Big) - \frac{\kappa^{2}}{2}S_{k,z}S_{k+1,z}\right]|\vec n\rangle
\end{eqnarray}
which has a structure analogous to the usual XXZ spin chain \cite{Lamers:2015dfa}
\begin{equation}
H_{XXZ}= \frac{\lambda}{2\pi\, J}\sum_{k=1}^{L}\Big[\frac{1}{4}- S_{k,\,x}S_{k+1,\,x}-S_{k,\,y}S_{k+1,\,y}- \Delta S_{k,\,z}S_{k+1,\,z}\Big].
\end{equation}
Comparing the two we get, 
\begin{equation}
 \Delta=1+\frac{\kappa^2 J^2}{2\,\lambda} %+\mathcal{O}(\kappa^4),
 \end{equation}
  i.e the `anisotropy' parameter $\Delta >1$ always in the case of our deformed spin chain. Comparing with the known structure of anisotropic XXZ spin chain, one can say that the above system is in the Neel or the anti-ferromagnetic phase. One more thing to note here is that usually in term of the quantum group parameter $q$, the anisotropy is written as
\begin{equation}
\Delta  = \frac{q+q^{-1}}{2}.
\end{equation}
One can now use the relation $q = \text{exp}\Big[ -\frac{2\eta}{g(1+\eta^2)}  \Big]=\text{exp}\Big[ -\frac{\kappa}{g\sqrt{1+\kappa^2}}  \Big] $ and expand in small $\kappa$ to find exact agreement with the expression of $\Delta$ found  the coherent state spin chain provided we identify $g = \frac{\lambda}{J^2}$ as the effective coupling.
%%%%%%%%%%%%%%%%%%%%%%%%%%%%%%%%%%%%%%%%%%%%%%%
\subsection{Circular string solutions}
Now we will move to studying a particular class of classical circular string solutions  the perspective of the anisotropic $ SU(2) $ spin chain Hamiltonian constructed in the previous section. These kind of solutions were first demonstrated in \cite{Kruczenski:2003gt} and were further detailed in \cite{Dimov:2004qv, Ryang:2004pu}. We vary the sigma model (\ref{e17}) w.r.t $\Phi$ and $\Theta$ in order to obtain the equations of motion,
\begin{eqnarray}
-2\dot{\Theta}\sin\Theta + \frac{\lambda}{J^2}\partial_{\sigma}(\Phi'\sin^2\Theta) &=& 0 \nonumber\\
\frac{\lambda}{J^2}\Theta'' + 2\sin\Theta\dot{\Phi}- \frac{\lambda}{J^2}\sin\Theta\cos\Theta \Phi'^2 - \kappa^2 \sin\Theta\cos\Theta &=& 0.
\end{eqnarray}

While the first equation is identical to that of \cite{Kruczenski:2003gt}, the second equation gets modified due to the presence of background deformations. One can assume that the boundary conditions on the spin chain remains unaltered and consider the deformation as a perturbation on an otherwise closed periodic chain namely,
\begin{equation}
\Phi(\sigma+L)=\Phi(\sigma).
\end{equation}

Hence one consistent ansatz is to consider $\partial_{\sigma}\Phi = 0$, which  using the first equation immediately gives $\partial_t \Theta =0$, and this in turn  the second equation indicates $\partial_{t}^2\Phi = 0$. This is of course only one branch of the solution. We will shortly consider the other possibilities also. In this case however we can assume that the associated classical string solution has a spinning ansatz of the form $\Phi = \omega\tau$. Then we are lead to only one effective equation of motion
\begin{equation}
\Theta '' + \frac{2\omega}{\tilde\lambda}\sin\Theta-\frac{\kappa^2}{\tilde\lambda}\sin\Theta\cos\Theta = 0,
\end{equation}
where we have redefined the effective coupling $\frac{\lambda}{J^2} = \tilde\lambda$. Comparing with the Jacobi differential equations, one could easily write an exact solution  in $\kappa$ for the string,
\begin{equation}
\tan\Theta(\s) = \mathbf{sn}\Big[ \sqrt{\frac{2\omega-\kappa^2}{2\tilde{\lambda}}}~\s ~|~ \frac{2\omega+\kappa^2}{2\omega-\kappa^2}   \Big].
\end{equation}
Here, we use the usual elliptic Jacobi functions and the boundary condition $\tan\theta(\s = 0) = 0$. This explicitly brings out the periodic nature of the solution along $\s$. But since we are more interested in the conserved quantities of motion, we will start with
\begin{equation}
\Theta' = \pm \left[ a + b \cos\Theta + c \cos^2\Theta    \right]^{1/2},~~b = \frac{4\omega}{\tilde\lambda},~c =- \frac{\kappa^2}{\tilde\lambda}.
\end{equation}
Here $a$ is a constant of integration. The above equation of motion looks like that of a particle in a trigonometric potential. Since the effective coupling is fixed, we can always put $c$ to be small. So now there are two cases to be considered, either $a>b>c$ or $b>a>c$. For our case, we will consider the former for simplicity. We can now write the integrals corresponding to the conserved charges associated with the spin chain:
\begin{eqnarray}
J &=& \int d\sigma = 4\int_{0}^{\Theta_0} \frac{d\Theta}{\sqrt{a+b\cos\Theta+c \cos^2\Theta}}\nonumber\\
J_{\Phi} &=& S_z = -\frac{1}{2} \int  \cos\Theta~ d\sigma= -2\int_{0}^{\Theta_0} \frac{\cos\Theta~d\Theta}{\sqrt{a+b\cos\Theta+c \cos^2\Theta}}.
\end{eqnarray}

We can easily integrate and find out the charges, albeit noting that $c$ is small we can expand the expressions upto first order in $c$ ,
\begin{eqnarray}
J &=& \frac{8}{\sqrt{a+b}}\mathbb{K}(x)+\frac{2c}{\sqrt{a+b}(a-b)b^2}\Big[ (b^2-2 a^2)\mathbb{E}(x) -2a(a-b)\mathbb{K}(x)   \Big], \nonumber\\ 
S_z &=& -\frac{4}{\sqrt{a+b}}\Big[ (a+b)\mathbb{E}(x)-a\mathbb{K}(x)\Big] \nonumber\\ 
&+& \frac{2c}{3b^3\sqrt{a+b}(a-b)}\Big[(8a^3 - 5ab^2)\mathbb{E}(x)-(a-b)(8a^2+b^2)\mathbb{K}(x)    \Big] ,\nonumber\\ 
x &=& \frac{2b}{a+b}.
\end{eqnarray}

Therefore, the charges could be schematically expanded as, $Q = Q^{(0)}+c Q^{(1)}$. At this stage, it is noteworthy to mention that for, $c=0$ our solutions are little different from the ones obtained in \cite{Kruczenski:2003gt}. This is due to the fact that here we have used a different inequality relation between the constants. Here $\mathbb{E}$ and $\mathbb{K}$ are the usual complete elliptic integrals. 

Our next task would be to calculate the total energy associated with the stringy configuration where we remind ourselves that the total Hamiltonian with $\Phi' = 0$ is given by
\begin{equation}
H = \frac{1}{2}\Big[\tilde{\lambda} \Theta'^2 + \kappa^2 \sin^2\Theta    \Big].
\end{equation}

Finally, we can write down the energy integral as follows,
\begin{eqnarray}
\gamma = E &=&  \frac{\tilde{\lambda}}{2}\int_{0}^{\Theta_0} \frac{(\Theta'^2 +\frac{\kappa^2}{\tilde{\lambda}}\sin^2\Theta)~d\Theta}{\sqrt{a+b\cos\Theta+c \cos^2\Theta}},\\ \nonumber
&=& \frac{\tilde{\lambda}}{8}\Big[aJ - 2b S_z   \Big]+\frac{\tilde{\lambda}c}{2}\int_{0}^{\Theta_0} \frac{\cos 2\Theta~d\Theta}{\sqrt{a+b\cos\Theta+c \cos^2\Theta}}.\\ \nonumber
&=& E^{(0)}+\tilde{\lambda}c E^{(1)}. \nonumber
\end{eqnarray}

It is therefore straightforward to find out the correction term to energy, which, as we have seen  the Hamiltonian is of $\mathcal{O}(\kappa^2)$. This is evident that this correction also bears contribution  $J$ and $S_z$ and the undeformed charges obey the same dispersion relation as in \cite{Kruczenski:2003gt}. Therefore, we can write,
\begin{equation}
E^{(1)} =  \frac{1}{8\sqrt{a+b}(a-b)b^2}\Big[\left(5 a b^2-2 a^3\right) \mathbb{E}(x)+2(a-b) \left( a^2-2 b^2\right) \mathbb{K}(x)\Big]
\end{equation}
where $x$ is the same as we had defined before. 

There is of course another set of non-trivial solutions where we can have $\partial_{\sigma}\Phi \neq 0$. In this case, we could still assume that $\Theta$ is a function of $\s$ only, so that the $\Phi$ equation of motion gives,
\begin{equation}\label{phiprime}
\Phi' = \frac{A}{\sin^2\Theta},
\end{equation}
where $A$ is  a constant. Using the above, the $\Theta$ equation can be written in the form,
\begin{equation}
\tilde{\lambda}\Theta'' + 2\dot\Phi\sin\theta-\tilde{\lambda}\frac{A\cos\Theta}{\sin^3\Theta}-\kappa^2\sin\theta\cos\theta=0.
\end{equation}
It is clear here that one can take an ansatz of the form $\Phi = \omega\tau + \tilde\Phi(\s)$, where $\tilde\Phi(\s)$ satisfies (\ref{phiprime}). This immediately boils down to the fact that $\ddot\Phi=0$  the $\Theta$ equation. Integrating the above equation of motion, we can write it in the form,
\begin{equation}
\Theta'^2+\frac{A}{\sin^2\Theta}-\frac{4\omega}{\tilde{\lambda}}\cos\Theta +\kappa^2\cos^2\Theta =B,
\end{equation}
where B is another integration constant.  By a substitution of $x = \cos\Theta$ we can right it in a `particle in a potential' form,
\begin{equation}
x'^2 = ax^4+bx^3+cx^2+dx+e
\end{equation}
with $a=\kappa^2,~b=-d=-\frac{4\omega}{\tilde{\lambda}},~c=-B-\kappa^2,~\text{and}~e=B-A$. This is indeed a involved dynamical system if one tries to solve it in full. A simpler subsector of solutions without expanding around small $\kappa$ can be generated by choosing $e=0$, in this case, without going into much details, we can write down a solution for the circular string in the following form,
\begin{equation}
x=\cos\Theta(\s)=\frac{\gamma~\textbf{dn}^2\Big[\frac{\sqrt{\gamma(\alpha-\beta)}}{2} \kappa\s~|~\frac{\alpha(\gamma-\beta)}{\gamma(\alpha-\beta)}  \Big]}{1-\frac{\gamma-\beta}{\alpha-\beta}~\textbf{sn}^2\Big[\frac{\sqrt{\gamma(\alpha-\beta)}}{2} \kappa\s~|~\frac{\alpha(\gamma-\beta)}{\gamma(\alpha-\beta)}\Big]}
\end{equation}
Here $\textbf{sn}$ and $\textbf{dn}$ are usual Jacobi elliptic functions, while $\a, \b, \g$ are roots of the polynomial $g(x) = x^3+\frac{b}{a}x^2+\frac{c}{a}x-b$, with $b,c$  as defined earlier.

%%%%%%%%%%%%%%%%%%%%%%%%%%%%%%%%%%%%%
\section{Fast spinning strings on $(R\times S^5)_{\kappa}$}

It is natural to consider larger subsectors of the known theory and try to predict deformed spin-chain structures in connection with the case elaborated in the last section. We shall now try to probe the $SU(3)$ case for the deformed theory along the lines of  \cite{Hernandez:2004uw}.

\subsection{The background}

We start with the full deformed metric for the $\kappa$ deformed $AdS^5\times S^5$ \cite{Arutyunov:2013ega},

\begin{eqnarray}\label{eq:metrc-etaAdS5S5-sph-coord}
\begin{aligned}
	ds^2_{(\text{AdS}_5)_{\kappa}}=&-\frac{1+\rho^2}{1-\kappa^2\rho^2}dt^2
	+\frac{d\rho^2}{ \left(1+\rho^2\right)(1-\kappa^2\rho^2)}\\
	&  + \frac{\rho^2}{1+\kappa^2\rho^4\sin^2\zeta}\left( d\zeta^2+\cos ^2\zeta \, d\psi_1^2\right) 
	+\rho^2 \sin^2\zeta\, d\psi_2^2\,,
	\\
	\\
	ds^2_{(\text{S}^5)_{\kappa}}=&\frac{1-r^2}{1+\kappa^2 r^2}d\phi^2
	+\frac{dr^2}{ \left(1-r^2\right)(1+\kappa^2 r^2)}\\
	&  + \frac{r^2}{1+\kappa^2r^4\sin^2\xi}\left( d\xi^2+\cos ^2\xi \, d\phi_1^2\right) 
	+r^2 \sin^2\xi\, d\phi_2^2\,.
\end{aligned}
\end{eqnarray}
Also we have the $B$-fields $B=\frac{1}{2} B_{MN}\ dX^M\wedge dX^N$~\cite{Arutyunov:2013ega}
\begin{eqnarray}\label{eq:B-field-etaAdS5S5-sph-coord}
\begin{aligned}
	\widetilde{B}_{(\text{AdS}_5)_{\kappa}} &= +\frac{\kappa}{2} \left( \frac{\rho^4 \sin (2\zeta)}{1+\kappa^2 \rho^4\sin^2 \zeta} d\psi_1\wedge d\zeta + \frac{2 \rho}{1-\kappa^2 \rho^2}dt\wedge d\rho\right),
	\\
	\widetilde{B}_{(\text{S}_5)_{\kappa}} &= -\frac{\kappa}{2} \left( \frac{r^4 \sin (2\xi)}{1+\kappa^2 r^4\sin^2 \xi}d\phi_1\wedge d\xi + \frac{2r}{1+\kappa^2 r^2}d\phi\wedge dr\right).
\end{aligned}
\end{eqnarray}
It is easy to see that the contributions of the components $B_{t\rho}$ and $B_{\phi r}$ to the Lagrangian are total derivatives, and hence can be ignored.

It is worthwhile to note that the $(AdS)_{\eta}$ contains a singularity, but we won't be bothered with that part in the present analysis. We can put in $\rho = 0$ and $r= \cos\theta$ and perform the redefinition of the coordinates $\phi \to \phi_3$ and $\xi \to \psi$ to write the metric of $(\mathbb{R}\times S^5)_{\kappa}$ in the following form, 
 \begin{eqnarray}\label{newmetric}
\begin{aligned}
	\\
	ds^2_{(\mathbb{R}\times S^5)_{\kappa}}=&-dt^2+\frac{\sin^2\theta}{1+\kappa^2 \cos^2\theta}d\phi_3^2
	+\frac{d \theta^2}{1+\kappa^2 \cos^2\theta}\\
	&  + \frac{\cos^2\theta}{1+\kappa^2 \cos^4\theta\sin^2\psi}\left( d\psi^2+\cos ^2\psi \, d\phi_1^2\right) 
	+\cos^2\theta \sin^2\psi\, d\phi_2^2\,.
\end{aligned}
\end{eqnarray}
And also the single surviving component of NS-NS flux takes the form as,
\begin{eqnarray}\label{Bfield}
\begin{aligned}
	\\
	\widetilde{B}_{(\mathbb{R}\times S^5)_{\kappa}} &= -\frac{\kappa}{2} \left( \frac{\cos^4\theta \sin (2\psi)}{1+\kappa^2 \cos^4\theta\sin^2 \psi}d\phi_1\wedge d\psi \right).
\end{aligned}
\end{eqnarray}

\subsection{Constructing the spin chain}
Now to study spinning string solutions
in this background, we use the Polyakov action coupled to an antisymmetric B-field,
\begin{eqnarray}
S &=&\int d\sigma d\tau~(\mathcal{L}_G +\mathcal{L}_B)\\ \nonumber
&=&-\frac{\sqrt{\hat\lambda}}{4\pi}\int d\sigma d\tau
[\sqrt{-\gamma}\gamma^{\alpha \beta}g_{MN}\partial_{\alpha} X^M
\partial_{\beta}X^N - \epsilon^{\alpha \beta}\partial_{\alpha} X^M
\partial_{\beta}X^N B_{MN}] \ ,
\end{eqnarray}
where  $\sqrt{\hat\lambda}$ is the changed 't Hooft coupling as been described before,
$\gamma^{\alpha \beta}$ is the worldsheet metric and $\epsilon^{\alpha
	\beta}$ is the antisymmetric tensor defined as $\epsilon^{\tau
	\sigma}=-\epsilon^{\sigma \tau}=1$.

We now follow the analysis of \cite{Hernandez:2004uw} and start with the spinning string ansatz,
\begin{equation}
t=\xi\tau,~~\phi_1= \alpha + t +\varphi,~~\phi_2 = \alpha + t -\varphi,~~\phi_3 = \alpha +t + \phi.
\end{equation}
The $\xi$ here is just a constant parameter, not to be confused with the coordinate we used earlier. The metric then takes the following form,
\begin{eqnarray}
ds^{2}_{(R\times S^{5})_{\kappa}}&=&-\xi^{2}d \tau^{2}+g_{\theta\theta} d\theta^{2}+g_{\psi\psi} d\psi^{2}+g_{\phi_3\phi_3}( \xi d\tau +d\alpha+d\phi)^{2}\\ \nonumber &+&g_{\phi_1\phi_1}( \xi d\tau +d\alpha+d\varphi)^{2}+g_{\phi_2\phi_2}( \xi d\tau +d\alpha-d\varphi)^{2}.
\end{eqnarray}

Using conformal gauge we can then write down the metric part $\mathcal{L}_G$ of the Lagrangian for the bosonic spinning string in the usual way as we had done before. To take the large spin limit, we again remind ourselves that $\xi$ is large and we can let go of the terms containing $\dot X^{\mu}$, provided $\xi\dot{X^\mu}$ is finite. Taking the limit carefully, we arrive at the following expression,
\begin{eqnarray}
\mathcal{L}_G &=& -\frac{\sqrt{\hat\lambda}}{4\pi}\bigg[  {\theta'}^2 g_{\theta\theta} -\xi^2 +(\alpha'^2+ \xi^2-2\xi\dot\alpha)(g_{\phi_1\phi_1}+g_{\phi_2\phi_2}+g_{\phi_3\phi_3})\\ \nonumber &-& 2\xi(g_{\phi_1\phi_1}\dot\varphi+g_{\phi_3\phi_3}\dot\phi-g_{\phi_2\phi_2}\dot\varphi)+g_{\psi\psi}\psi'^2+g_{\phi_3\phi_3}\phi'^2 +(g_{\phi_1\phi_1}+g_{\phi_2\phi_2})\varphi'^2  \\ \nonumber &+& 2\alpha'\varphi' (g_{\phi_1\phi_1} - g_{\phi_2\phi_2}) +2\alpha'\phi' g_{\phi_3\phi_3}\bigg]
\end{eqnarray}
We can see  the form of the form of the lagrangian that at the limit $\xi\to\infty$ we have to make sure that $\xi^2\kappa^2$ remains a finite combination, which translates to the condition that $\kappa\to 0$. Dropping out the terms solely of  $\mathcal{O}(\kappa^2)$, throughout without any loss of generality, we can get the BMN string lagrangian,
\begin{eqnarray}
\mathcal{L}_G = \frac{\sqrt{\lambda}}{4\pi}\bigg[ 2\xi\dot\alpha &+& 2\xi\cos^2\theta\cos(2\psi) \dot\varphi+2\xi\sin^2\theta\dot\phi - \alpha'^2-\theta'^2 -\cos^2\theta\varphi'^2 \\ \nonumber   &-& \sin^2\theta\phi'^2-\cos^2\theta\psi'^2 -2\sin^2\theta\alpha'\phi' -2\cos^2\theta\cos(2\psi)\alpha'\varphi'   \\ \nonumber  &-& \xi^2\kappa^2\cos^2\theta \big( \sin^2\theta+\sin^2\psi\cos^2\psi\cos^4\theta   \big) \bigg]
\end{eqnarray}
Looking at the Lagrangian, we can easily spot that this is the same one derived in \cite{Hernandez:2004uw} albeit with new $\mathcal{O}(\sqrt{\lambda}\xi^2\kappa^2)$ terms added due to the deformation. We can now go back and discuss about the NS-NS flux contribution to the Lagrangian. One might naively note that the total Lagrangian starts at the $\mathcal{O}(\sqrt{\lambda}\kappa)$, which in our case can be neglected in the BMN limit. But there is a subtlety here. Putting in the coordinate transformations, the B-field finally has the form,
\begin{equation}
B= -\frac{\kappa}{2} \frac{\cos^4\theta \sin (2\psi)}{(1+\kappa^2 \cos^4\theta\sin^2 \psi)}\left[(d\alpha+ dt+d\varphi)\wedge d\psi \right]
\end{equation}
Under the BMN limit, we can see one of the flux terms will survive, so we could write
\begin{equation}
\mathcal{L}_B = -\frac{\sqrt{\lambda}\kappa}{4\pi} (\cos^4\theta \sin (2\psi))~\dot t \psi'+\mathcal{O}(\sqrt{\lambda}\xi\kappa^3)
\end{equation}
We can ignore other terms since they will have $\dot X$ contribution. We should note here that $\dot t = \xi$ and of course $\mathcal{O}(\sqrt{\lambda}\xi\kappa)$ terms give finite contribution in the required limit.

As in the case before, the Virasoro constraint $T_{\tau\sigma} = 0$ gives in the leading order of $\kappa$,
\begin{equation}
2\xi\alpha' + 2\xi\phi'\sin^2\theta+2\xi\varphi'\cos^2\theta\cos(2\psi) = 0
\end{equation}
Putting the value of $\alpha'$ back into the Lagrangian and resorting to target space time coordinate $t=\xi\tau$, we can write the total action in the following suggestive form,

\begin{eqnarray}\label{kappa3}
S&=& \frac{\sqrt{\lambda}\xi}{2\pi}\int d\sigma dt ~\bigg[ \dot\alpha  +\sin^2\theta\dot\phi+\cos^2\theta\cos(2\psi)\dot\varphi \\ \nonumber&-&\frac{\kappa^2}{2}\cos^2\theta \big( \sin^2\theta+\sin^2\psi\cos^2\psi\cos^4\theta)    \bigg] \\  \nonumber
&-&\frac{\sqrt{\lambda}}{4\pi\xi}\int d\sigma dt ~\bigg[ \theta'^2+\cos^2\theta(\psi'^2+\sin^2(2\psi)\varphi'^2)+\sin^2\theta\cos^2\theta (\phi' - \cos(2\psi)\varphi')^2 \bigg]\\ \nonumber &-&\frac{\sqrt{\lambda}\kappa}{4\pi \xi}\int d\sigma dt~[\cos^4\theta \sin (2\psi)]\psi' .\nonumber
\end{eqnarray}
Here, we can see that the momentum associated to $\alpha$ is simply $P_\alpha =\sqrt{\lambda}\xi $, which we can identify with the spin chain conserved charge $J$. Then we can 
easily write the total Hamiltonian in the form,
\begin{eqnarray} \label{effect}
\mathcal{H}_\kappa &=& \frac{\lambda}{4\pi J}\int d\sigma ~\bigg[ \theta'^2+\cos^2\theta(\psi'^2+\sin^2(2\psi)\varphi'^2)\\ \nonumber &+&\frac{1}{4}\sin^2 (2\theta) (\phi' - \cos(2\psi)\varphi')^2 \bigg] +\frac{J}{2\pi}\int d\sigma ~\bigg[ \frac{\kappa^2}{2}\cos^2\theta \big( \sin^2\theta+\sin^2\psi\cos^2\psi\cos^4\theta)       \bigg] \\ \nonumber
&+& \frac{\lambda\kappa}{4\pi J}\int d\sigma ~[\cos^4\theta \sin (2\psi)]\psi' \\ \nonumber
&=& \mathcal{H}_{SU(3)}+\mathcal{H}_{D}+\mathcal{H}_{B}.
\end{eqnarray}

Here $\mathcal{H}_{SU(3)}$ corresponds to the Hamiltonian of usual $SU(3)$ spin chain in the continuum limit, also $\mathcal{H}_D$ and $\mathcal{H}_B$ are the deformation term and the NS-NS flux contribution respectively. We can note here that the usual $SU(3)$ spin chain has an effective coupling $\frac{\lambda}{J}$, while the NS-NS contribution has an effective coupling $\frac{\lambda\kappa}{J}$, so that we can treat it as an addition of a small interaction to the spin chain. The contribution due to the deformation, on the other hand, occurs at $\mathcal{O}(\kappa^2)$  as was in the case of $SU(2)$. One can easily recover the results of $SU(2)$ case (up to some re-scalings) by simply putting in $\psi = 0$.\par

\subsection{Deformed spin chain in terms of coherent states}
Next we try to match this Hamiltonian (\ref{effect}) as a continuum limit of $SU(3)$ Heisenberg type model, albeit with added corrections. We proceed pretty much along the lines similar to the SU(2) case. Here, we start with the $SU(3)$ coherent spin state as described in \cite{Hernandez:2004uw}. It must be noted here that there is no a-priori reason to expect that for the deformed  symmetry the usual coherent state will still be enough to capture the underlying spin chain, although it is indeed interesting to see how the predicted and actual terms deviate from each other in this case. For the action of the (unbroken) symmetry group on the $k^{th}$ site we get,
\begin{equation} \label{cohSU3}
|\vec n_k\rangle=\cos \theta_k \,\cos \psi_k e^{i\,\varphi_k}|1\rangle+\cos \theta_k \,\sin \psi_k e^{-i\,\varphi_k}|2\rangle+\sin \theta_k \,e^{i\,\phi_k}|3\rangle.
\end{equation}
With $0\leq\theta<\pi,~0\leq\psi<2\pi,~0\leq\phi_i<2\pi$. The coherent state for the full spin chain is again,
\begin{equation}
|\vec n \rangle=\prod_{k=1}^{L}|\vec n_{k}\rangle.
\end{equation}
For the  $\kappa=0$ part we get back the undeformed SU(3) Heisenberg spin-chain,
\begin{equation}
\mathcal{H}_{\kappa=0}=\frac{\lambda}{2\pi J}\sum_{k=1}^{J}\langle\vec n | \Big[\frac{4}{3}-\sum_{a=1}^{8}\lambda_{k,\,a}\lambda_{k+1,\,a}\Big]|\vec n\rangle,
\end{equation}
which gives the appropriate terms i.e. $\mathcal{H}_{SU(3)}$ in $L\rightarrow \infty$ limit. Here, $\lambda_{a}'s$
are the usual Gell-Mann matrices,$~a=1,...8$. Details of our convention and action of coherent state on these matrices are given in the appendix A. Next we try to generate the other terms at the order of of $\kappa$ and $\kappa^2$  this coherent state itself. To do so we start with the following general linear combination,
\begin{align}
\begin{split} \label{lin}
I=\sum_{a_b=1}^{8}a_b\langle n_{k}|\lambda_{k,b}|n_k\rangle \langle n_{k+1}|\lambda_{k+1,b}|n_{k+1}\rangle.
\end{split} 
\end{align}
Now taking the continuum limit, we get, 
\begin{align}
\begin{split}
I=&L\int d\sigma \Bigg[\frac{a_{8}}{12}+
\cos ^4\theta  \left(\sin ^2(2 \psi ) \left(a_{2} \sin ^2(2 \varphi )+a_{1} \cos ^2(2 \varphi )\right)+a_{3} \cos ^2(2 \psi )\right)\\&+\sin ^2(2 \theta ) \Big(\cos ^2\psi  \left(a_{4} \cos ^2(\varphi -\phi )+a_{5} \sin ^2(\varphi -\phi )\right)\\&+\sin ^2\psi  \left(a_{6} \cos ^2(\varphi +\phi )+a_{7} \sin ^2(\varphi +\phi )\right)\Big)+\frac{1}{4} a_{8} \cos (2 \theta ) (3 \cos (2 \theta )-2)\\&+\frac{\phi'}{2\,L}\Big(\sin ^2(2 \theta ) \left((a_{5}-a_{4}) \cos ^2\psi  \sin 2( \varphi - \phi )+(a_{6}-a_{7}) \sin ^2\psi  \sin (2 (\varphi +\phi ))\right)\Big)\\&
+\frac{\varphi'}{2\,L}\Big(2 \sin (4 \varphi ) (a_{1}-a_{2}) \cos ^4\theta  \sin ^2(2 \psi )+\sin ^2(2 \theta )\Big((a_{4}-a_{5}) \cos ^2\psi  \sin 2( \varphi - \phi )\\&+(a_{6}-a_{7}) \sin ^2\psi  \sin (2 (\varphi +\phi ))\Big)\Big)+\frac{\psi'}{L}\Big(\cos ^4\theta  \sin (4 \psi ) \Big(-a_{2} \sin ^2(2 \varphi )\\&+a_{3}-a_{1} \cos ^2(2 \alpha )\Big)+\sin ^2(2 \theta ) \sin \psi  \cos \psi  \Big(a_{4} \cos ^2(\varphi -\phi )+a_{5} \sin ^2(\varphi -\phi )\\&-a_{6} \cos ^2(\varphi +\phi )-a_{7} \sin ^2(\varphi  +\phi )\Big)\Big)+\frac{\theta'}{4\,L}\Big(4 \sin (\theta ) \cos ^3(\theta ) \Big(\sin ^2(2 \psi ) (\cos (4 \alpha ) \\&(a_{1}-a_{2})+a_{1}+a_{2})+a_{3} \cos (4 \psi )+a_{3}\Big)+\sin (4 \theta ) \Big(-2 \cos ^2(\psi ) ((a_{4}-a_{5}) \cos (2 \varphi -2 \phi )\\&+a_{4}+a_{5})-4 \sin ^2(\psi ) \left(a_{6} \cos ^2(\varphi +\phi )+a_{7} \sin ^2(\varphi +\phi )\right)+3 a_{8}\Big)-2 a_{8} \sin (2 \theta )\Big)\Bigg].
\end{split}
\end{align}
Here we have kept terms upto linear derivative and neglected the quadratic derivative term in the large $L$ limit.  Now we focus first how to generate the term coming   $B$ field contribution.  We then pick out the $\psi' $ term in the expansion,
\begin{align}
\begin{split} \label{eqt}
\int d\sigma &\Big[\cos ^4\theta  \sin (4 \psi ) \left(-a_{2} \sin ^2(2 \varphi )+a_{3}-a_{1} \cos ^2(2 \varphi )\right)+2(\cos^2\theta-\cos^4\theta)\sin (2\psi )\\& \left(a_{4} \cos ^2(\varphi -\phi )+a_{5} \sin ^2(\varphi -\phi )-a_{6} \cos ^2(\varphi +\phi )-a_{7} \sin ^2(\varphi +\phi )\right)\Big]\psi',
\end{split}
\end{align}
From (\ref{eqt}) it is evident we cannot generate only the B field term, as a by-product, other terms will naturally be there. We rescale all the coefficients $a_{i}$ by a factor of $\frac{\lambda \kappa}{4\pi J}$ to make connection with the spin chain  the sigma model. Then we can set $ a_{4}=a_{5}, \text{and} ~a_{6}=a_{7} .$ This will give,
\begin{align}
\begin{split}
\frac{\lambda \kappa}{4\pi J}\int d\sigma &\Big[\cos ^4\theta \sin (4 \psi ) \left(-a_{2} \sin ^2(2 \varphi )+a_{3}-a_{1} \cos ^2(2 \varphi )\right)\\&+2(a_4-a_6)(\cos^2\theta-\cos^4\theta)\sin (2\psi )\Big]\psi',
\end{split}
\end{align}
Then we can set $a_1=a_2=a_3$, leaving us with ,
\begin{align}
\begin{split}
\frac{\lambda \kappa}{4\pi J}\int d\sigma &\Big[2(a_4-a_6)(\cos^2\theta-\cos^4\theta)\sin (2\psi )\Big]\psi',
\end{split}
\end{align}
So we will have these two term associated with $\psi'.$ Not only that, we still have the following extra terms also,
\begin{align}
\begin{split}
&\frac{\lambda \kappa}{8\pi J}\int d\sigma\Big[ \sin (2 \theta ) (\cos (2 \theta ) (a_{3}+2 (a_{6}-a_{4}) \cos (2 \psi )-2 (a_{4}+a_{6})+3 a_{8})+a_{3}-a_{8})\Big]\theta'+\\&\frac{\lambda \kappa}{96\pi J} \int d\sigma \Big[ 3 \cos (4 \theta ) (a_{3}-2 (a_{4}+a_{6})+3 a_{8})+12 (a_{3}-a_{8}) \cos (2 \theta )+9 a_{3}\\&+12 (a_{4}-a_{6}) \sin ^2(2 \theta ) \cos (2 \psi )+6 (a_{4}+a_{6})+11 a_{8}\Big].
\end{split}
\end{align}
Also it is evident that these extra terms cannot be set to zero entirely.  At best we can choose, 
\begin{equation}
a_3=2 a_4+2 a_6-3 a_8,\,\, a_8=\frac{a_4+a_6}{2}.
\end{equation}
Then we will be left with,
\begin{equation} \label{eqt1}
\frac{\lambda \kappa}{8\pi J}\int d\sigma \Big[ \frac{1}{3} \left(3 (a_{4}-a_{6}) \sin ^2(2 \theta ) \cos (2 \psi )+4 (a_{4}+a_{6})\right)-(a_{4}-a_{6}) \sin (4 \theta ) \cos (2 \psi )\theta'\Big].
\end{equation}
 (\ref{eqt1}) we here we can choose,
\begin{equation}a_4=\pm 1+a_6, ~a_6=\mp\frac{1}{2}.
\end{equation}
So we can't reproduce our B-field term exactly in this procedure, due to the presence of extra contributions which can't be accounted for via the $SU(3)$ symmetry of the coherent state. So we end up with spin-chain representation for only a part of the B-field,  
\begin{align}
\begin{split} \label{extra}
\frac{\lambda\kappa}{4\pi J}\int d\sigma ~[\cos^4\theta \sin (2\psi)]\psi' &=\pm \frac{\lambda (1-\beta)}{2\pi J}\lim_{L\rightarrow \infty} \sum_{k=1}^{L}\sum_{a_b=4,5,6,7}\langle n_{k}|\lambda_{k,b}|n_k\rangle \langle n_{k+1}|\lambda_{k+1,b}|n_{k+1}\rangle\\&+\frac{\lambda\kappa}{4\pi J}\Big[\int d\sigma \Big(2\cos^2\theta\sin (2\psi )\psi'-\frac{1}{4}  \sin (4 \theta ) \cos (2 \psi )\theta'\\&+\frac{L}{4} \sin ^2(2 \theta ) \cos (2 \psi )\Big)\Big]
\end{split}
\end{align}
where we have $\beta=1+\frac{\kappa}{4}$.  

Next we try to reproduce the $\frac{\lambda \kappa^2}{J}$ terms. We start with the same linear combination (\ref{lin}). In previous case we kept linear derivative term as they are associated with $\frac{\lambda\kappa }{J}.$ But for this case we neglect them as they are multiplied with $J \kappa^2 .$ Then we can easily see that, 
%We have made the assumptions $\lambda \kappa^2 (\partial_{\sigma} X^{A}) $ can be neglected and but we retain terms like $\lambda \kappa (\partial_{\sigma} X^A).$ Here $ X^{A}=\{\phi, \psi,\theta,\varphi\}.$ 
\begin{align}
\begin{split} \label{extra1}
& \frac{J\,\kappa^2\, L }{4\pi}\int d\sigma ~\bigg[\cos^2\theta \big( \sin^2\theta+\sin^2\psi\cos^2\psi\cos^4\theta)       \bigg]\\&=\frac{\lambda\, (\alpha-1)}{2\pi J }\lim_{L\to\infty} \sum_{k=1}^{L}\sum_{b=1,2,4,5,6,7}\langle n_{k}|\lambda_{k,b}|n_k\rangle \langle n_{k+1}|\lambda_{k+1,b}|n_{k+1}\rangle\\&-\frac{J\,\kappa^2\,L}{4\pi}\int d\sigma ~\bigg[\cos^4\theta\sin^2\theta \sin^2\psi\cos^2\psi\Big]  ,
\end{split}
\end{align}
where $\alpha=1+\frac{\kappa^2 J^2}{8\,\lambda}$ is an anisotropy parameter. \par
%Of course there is another term in the effective hamiltonian (\ref{effect}),
%$$\frac{J}{2\pi}\int d\sigma \cos^4\theta\sin^2\theta \sin^2\psi\cos^2\psi.$$ 
From (\ref{extra}) and (\ref{extra1}) it is evident that are various extra terms left over that cannot be generated by probing the theory with $SU(3)$ coherent state. This is not surprising fact as $SU(3)$ symmetry of the Hamiltonian has been broken at the perturbative level.  One can hope to study these terms by starting with quantum deformed coherent state, writing the full Hamiltonian and then expanding it in terms of the deformation parameter. The problem of having an extra term at the quadratic order of the deformation parameter was addressed in \cite{Frolov:2005iq}  via a properly chosen non-unitary transformation on the spin chain and then taking the continuum limit. We leave these questions open for possible future studies. \par
As a final step, we collect the parts of the Hamiltonian that can be written in terms of the $SU(3)$ coherent state and write them as following,
\begin{align}
\begin{split}
\tilde H_{\kappa}=\frac{\lambda}{2\pi J}\sum_{k=1}^{L}\langle\vec n | \Big[&\frac{4}{3}-\sum_{a=3,8}\lambda_{k,\,a}\lambda_{k+1,\,a}-(1-\Delta_{1}^{\pm})\sum_{b=4,5,6,7}\lambda_{k,\,b}\lambda_{k+1,\,b}\\&-(1-\Delta_2)\sum_{c=1,2} \lambda_{k,\,c}\lambda_{k+1,\,c}\Big]|\vec n\rangle,
\end{split}
\end{align}
where, $\Delta_{1}^{\pm}= (\alpha\mp \beta-1\pm 1)$ and $\Delta_{2}=(\alpha-1)=\frac{\kappa^2J^2}{8\lambda}.$ So  there are two choices for the $\Delta_{1},$
\begin{equation}
\Delta_{1}^{+}=\frac{\kappa^2J^2}{8\lambda}- \frac{\kappa}{4},~\Delta_1^{-}=\frac{\kappa^2J^2}{8\lambda}+\frac{\kappa}{4}.
\end{equation}
 this is evident that there are two situations based on choice of the constants. Now  for the case of $\Delta_{1}^{+}$  we will have $1-\Delta_{1}^{+} >1$ and $1-\Delta_{2} <1$, so we would get two competing anisotropy parameters. On the other hand, for $ \Delta_{1}^{-}$ case we will have both $1-\Delta_{1}^{-}$ and $1-\Delta_{2}$ to be less than one.  These two regimes of different anisotropies offer a rich physical structure in the spin chain that we would like to explore in detail elsewhere.

\par
%%%%%%%%%%%%%%%%%%%%%%%%%%%%%%%%%%%%%%%%%%%%%%%%%%%%%
\subsection{Circular string solutions}
Now, we can discuss the circular string spectrum for the deformed $SU(3)$ model. Let us start by writing the equations of motion for the deformed action [\ref{kappa3}],
\begin{eqnarray}
&&\sin(2\theta)\dot\theta=\frac{\lambda}{4J^2}\partial_{\s}\left[ \sin^2(2\theta)(\phi'- \cos(2\psi)\varphi')  \right],\\ \nonumber
&&\partial_{t}(\cos^2\theta \cos(2\psi))=\frac{\lambda}{J^2}\partial_{\s}\left[ \cos^2\theta\sin^2(2\psi)\varphi' +\frac{1}{4}\sin^2(2\theta)  (\phi'- \cos(2\psi)\varphi') \cos(2\psi)  \right],\\ \nonumber
&&\cos^2\theta\sin(2\psi)\bigg[ \dot\varphi+\frac{\lambda}{J^2} \left(\cos^2\theta\cos(2\psi)\varphi'^2  +\sin^2\theta\varphi'\phi' \right)    \bigg]+\frac{\lambda\kappa}{2J^2}\cos^4\theta\cos(2\psi)\\
&+&\frac{\kappa^2}{8}\sin(4\psi)\cos^4\theta = \frac{\lambda}{2J^2}\partial_{\s}\left[ \cos^2\theta\psi'+\frac{\kappa}{2}\cos^4\theta\sin(2\psi)    \right].\\ \nonumber
\end{eqnarray}
Also we can write the complicated $\theta$ equation in the form
\begin{equation}
\frac{\lambda}{J^2}\theta'' + \mathcal{V}(\theta,\psi,\phi,\varphi, \kappa) = 0
\end{equation}
where we have,
\begin{eqnarray*}
\mathcal{V} &=& \sin (2\theta) \left[ \dot\phi- \cos(2\psi)\dot\varphi  +\frac{\lambda}{2J^2}\left[ \psi'^2 +\sin^2(2\psi)\varphi'^2- \cos(2\theta)(\phi'- \cos(2\psi)\varphi')^2  \right] \right] \\ \nonumber
&-& \frac{\kappa^2}{2}\left[  \sin(2\theta)\cos(2\theta)-6 \sin^2\psi\cos^2\psi \cos^5\theta \sin\theta    \right]+\frac{2\lambda\kappa}{ J^2}\left[   \cos^3\theta\sin\theta\cos(2\psi)   \right]
\end{eqnarray*}
Solving these equations of motion in full generality appears to be a herculean task. For simplicity, we will find the energy states analysed in \cite{Hernandez:2004uw} for the $SU(3)$ case. For more general string configurations  $SU(3)$ spin chains, see for example \cite{Kristjansen:2004ei}, \cite{Kristjansen:2004za}, \cite{Freyhult:2005fn}. Let us now consider a simplified solution where we can consistently employ the ansatz
\begin{equation}
\theta  = \theta_0,~\psi = \psi_0,~\varphi = m\s+g\tau,~ \phi = n\s+h\tau
\end{equation}
These $m,n$ are the winding numbers of the circular string and $g,h$ are constants depending on $m,n,$ and $\kappa$. Under this simplification, it is easy to write the conserved charges for the solution as following,
\begin{eqnarray}
P_{\varphi} &=& J \cos^2\theta_0 \cos 2\psi_0 = J_1 - J_2\\ \nonumber
P_{\phi} &=& J \sin^2\theta_0  = J_3\\ \nonumber
\end{eqnarray}
The Hamiltonian for this case simplifies to the form
\begin{eqnarray}
\mathcal{H}_\kappa &=& \frac{\lambda}{2 J}\ ~\bigg[ m^2\cos^2\theta_0\sin^2(2\psi_0)+\frac{n^2}{4}\sin^2 (2\theta_0)+\frac{m^2}{4}\sin^2 (2\theta_0)\cos^2(2\psi_0) \\ \nonumber &-& \frac{mn}{2}\sin^2(2\theta_0)\cos(2\psi_0) \bigg]+\bigg[ \frac{J\kappa^2}{2}\cos^2\theta_0 \big( \sin^2\theta_0+\sin^2\psi_0\cos^2\psi_0\cos^4\theta_0)       \bigg] \\ \nonumber
\end{eqnarray}
One can notice here that the $NS$ flux term does not contribute in this case. We then use the relations consistent with the definition of charges
\begin{equation}
\frac{J_1}{J} = \cos^2\theta_0\cos^2\psi_0,~\frac{J_2}{J} = \cos^2\theta_0\sin^2\psi_0,~\frac{J_3}{J} = \sin^2\theta_0,
\end{equation}
and write the total energy of the string as,
\begin{eqnarray}\label{su3E}
E_{\kappa} &=& \frac{\lambda}{2J} \frac{1}{J^2} \Big[  4m^2 J_1J_2+(m-n)^2 J_1J_3+(m+n)^2 J_2J_3\\ \nonumber  &+&
\frac{\kappa^2}{\tilde{\lambda}}\big(J_1J_2+J_2J_3+J_1J_3-\frac{J_1J_2J_3 }{J}\big)  \Big]\\ \nonumber
\end{eqnarray}
We can easily see that the energy reduces to the exact undeformed value when $\kappa = 0$ \cite{Hernandez:2004uw}. The curious thing to note here is that in the deformation term (of $\mathcal{O}(\kappa^2)$) the quadratic terms of $J_i$'s explicitly correspond to the terms in the deformed spin chain that can be written in terms of the $SU(3)$ coherent state. On the other hand the cubic term comes  the Hamiltonian contribution of $\sim \cos^4\theta\sin^2\theta \sin^2\psi\cos^2\psi\ $, which, as we have shown, can't be written in terms of the coherent state for $SU(3)$. This term explicitly points to the quantum group deformed symmetry which we can't capture in our analysis, at least by doing so from the undeformed symmetry considerations.

\section{Map to other deformed sigma models}
The new interest in YB deformed sigma models has taken its course along two different paths, namely the standard $q$-deformed theories and the Jordanian deformed theories. In both of these cases, the structure of classical $r$-matrices plays the central role in constructing the deformed sigma model. The point to stress here would be that the former is based on the modified Classical Yang-Baxter Equation (mCYBE), while the latter is simply based on the Classical Yang-Baxter Equation (CYBE). The sigma model on the squashed sphere and that of the $\eta$-deformed $AdS_5\times S^5$ superstring \cite{Delduc:2013qra} fall into the standard deformed category, while the Jordanian deformed theories have been studied in detail, for example, see \cite{Kawaguchi:2014qwa, Kawaguchi:2014fca} and references therein.

The standard $q$-deformed theories have been discussed in the literature for many years. At the spin chain level, the connection between fast-spinning string limits of squashed three-sphere and $(\mathbb{R}\times S^3)_{\eta}$ was shown in \cite{Kameyama:2014bua}. Sure enough, this connection has deep rooted indications for the theory. For example, it was beautifully shown in \cite{Hoare:2014pna}  that one could reproduce both the $(\mathbb{R}\times S^3)_{\eta}$  and the Squashed $S^3$ starting  the celebrated Fateev $O(4)$ model \cite{Fateev:1996ea} and putting in different values of the squashing parameters. The most general deformation at the quantum level is that of the low-energy limit of the Leigh-Strassler deformed \cite{Leigh:1995ep} one-loop spin-chain \cite{Bundzik:2005zg}, which for $SU(2)$ case can be written in the form as following,
\begin{equation}
\frac{\lambda}{16\pi|q|}\sum_{i=1}^{L}\Big[ \left(\frac{1+qq^*}{2}+hh^*\right) \mathbb{I}-\left(\frac{1+qq^*}{2}-hh^*\right)\s_{z,i}\s_{z,i+1}  -2q\s^{-}_{i}\s^{+}_{i+i}-2q^*\s^{+}_{i}\s^{-}_{i+i} \Big].
\end{equation}
Here $q$ and $h$ are the quantum group parameters and $\s^{\pm} = \s^x+i\s^y$. For example, the choice $q =\text{exp}[\beta_I - i\beta_R]$ and $h=0$ produces the general complex $\beta$ deformed spin chain model. One can explicitly see that if $\b_{R} = 0$, the fast spinning string limit  of the action corresponding to this geometry is given by,
\begin{equation}
\mathcal{S} = \mathcal{S}_{SU(2)}-\frac{\lambda}{16\pi L}\int d\tau d\sigma~ \b_{I}^2\sin^2 2\theta,
\end{equation}
which has  exactly the  same form as (\ref{actionsu2}).

This connection between the continuum spin chain limits of the  $\beta$ deformed models and $\eta$-deformed models also sustain for higher dimensional cases, albeit with some subtlety. One must mention here the work \cite{Matsumoto:2014nra}, where the authors generate $\g$-deformations of the $AdS_5\times S^5$ superstring as Yang-Baxter sigma models with classical r-matrices satisfying the classical Yang-Baxter equation. As an example, let us consider the total Lunin-Maldacena geometry with complex $\b$ deformation \footnote{See Appendix B for some details of the background.} parameter $\b=\beta_I - i\beta_R$. Again we will consider the case $\b_{R} = 0$ and the  subsector has been shown to preserve integrable motion of classical strings \cite{Frolov:2005ty}. However, in \cite{Puletti:2011hx}, the authors have discussed a rather unconventional limit of the background where $\b_I = 0$ and the quantum group parameter is taken as $q= \text{exp}(i\b)$, i.e. the purely imaginary deformation limit. The background  metric, in this case, is simply the real $\beta$ deformed metric multiplied by a conformal factor, however, the $NS$ fluxes are more involved. This adds to the fact that a particular deformed $SU(3)$ sector spanned by an anti-holomorphic field and two holomorphic fields (or vice versa) is supposed to be integrable for even imaginary $\beta$ \cite{Mansson:2007sh}. One must note here that only this does not guarantee the integrability of the whole model. In fact, the non-integrability of the purely imaginary $\b$ deformed background was argued via chaotic motion of strings in \cite{Giataganas:2013dha}.

For this imaginary $\beta$ case, the fast spinning limit has been considered in \cite{Puletti:2011hx}. In our parameterization of the undeformed coherent state, we can explicitly see that in addition to usual $SU(3)$ terms, the deformation term in their Hamiltonian has a form,
\begin{equation}
H_D\sim \frac{\beta_R^2}{2}\int d\sigma ~\bigg[\cos^2\theta \big( \sin^2\theta+\sin^2\psi\cos^2\psi\cos^4\theta)\bigg]
\end{equation}
which exactly has the similar form to our deformation term. Moreover, the $B$ field term in this case reads.
\begin{equation}
H_B \sim \b_R \Big[\sin\theta\cos\theta\cos^2\psi~\theta' + \sin^2\theta\cos^2\theta\cos\psi\sin\psi~\psi'     \Big]
\end{equation}
which seems to be different in our case. But this is intriguing that at least the level of metric, the action for the imaginary $\beta$ deformation reproduces same deformation term as is the case of $\eta$ deformed background in the fast spinning limit. Naively one could say that the string in this limit does not distinguish between different deformations and `perceives' only the leading order deformed geometry. The other intriguing observation is that in \cite{Puletti:2011hx} a spinning string solution for the $SU(3)$ sector has been performed and it has been explicitly shown that this energy can be reproduced via an algebraic Bethe equation. And this can be shown to exactly lead to the same correction to the energy quadratic in the deformation parameter as in (\ref{su3E}), even with the cubic term in $J_i$. For some details on this point, one can have a look at Appendix C.
 
 This is a rather new connection between these two drastically different theories, which is quite surprising since the $\beta$ deformed background is obtained by T dualities and S dualities on the target space, while the $\eta$ deformation is performed purely at the worldsheet sigma model level. It is a right moment to mention here that a recent analysis \cite{Roychowdhury:2017vdo} of spinning strings and Normal Variational Equations (NVEs) in the $\eta$ deformed $AdS_3$ and $S^3$ has puzzlingly found signs of non-integrability. Indeed the method of analysis is similar to that for non-integrability of imaginary $\b$ deformed case \cite{Giataganas:2013dha}. This remains a very mysterious point that one has to address with all its subtlety via better methods.
%%%%%%%%%%%%%%%%%%%%%%%%%%%%%%%%%%%%%%%
\section{Summary and final remarks}
We now summarize the key observations made in this paper. The purpose of the present work was to study the $ \eta $ deformed sigma models in the fast spinning limit and to explore possibilities whether these sigma models have any resemblance to that with the Heisenberg spin chain systems in its continuum limit. It turns out, rather expectedly, that in the fast spinning limit, a part of the string sigma model corresponding to $ (R \times S^{5})_{\eta} $ could be expressed as the continuum limit of $SU(3) $ Heisenberg-type spin chain with added anisotropic terms, provided we confine ourselves to nearest neighbor interactions. However, as a by-product of our analysis, we generate a few additional terms in the sigma model that eventually survive the BMN limit and cannot be expressed in terms of nearest neighbour interactions using the standard definition of $ SU(3) $ coherent states. Furthermore, as an interesting observation, we identify these additional contributions in the sigma model to be identical to that with the corresponding fast spinning strings with imaginary $ \beta $ deformations \cite{Puletti:2011hx}. The description of this  imaginary $ \beta $ model seems also to be plagued by the presence of terms cubic in coherent vectors of $SU(3)$. Our observation might actually point towards unveiling a new connection between two apparently drastically different looking string theories near the BMN limit. However, the reason why one is not able to match the B-field contributions in the two cases, and how it affects the analysis, eludes us so far.

Before we formally conclude, it is indeed noteworthy to mention that the fast spinning limit with fermionic excitations has never been explored in the context of $ q $ deformed sigma models.  It would be therefore an interesting exercise to extend the entire analysis for the fermionic sector (following, for example, \cite{Hernandez:2004kr}-\cite{Stefanski:2007dp}) and in particular to explore whether sigma models with YB deformations has any resemblance to that with the $ q $ deformed supersymmetric spin chains \cite{Arnaudon:1997ej}-\cite{Bazhanov:2008yc} in its continuum limit. This should be feasible in the case of $ \eta  $ deformation since it allows us to write down a Lagrangian description of the standard Green-Schwarz type \cite{Arutyunov:2015qva} that is quadratic in fermionic excitations. We hope to address this issue in near future.
\\ \\
%%%%%%%%%%%%%%%%%%%%%%%%%%%%%%%%%%%%%%%%%%%%%%%%%%%%%%%%%%%%%%%%%%%%%%

%%%%%%%%%%%%%%%%%%%%%%%%%%%%%%%%%%%%%%%%%%%%%%%%%%%%%%%%%%%%%%%%%%%%%%
{\bf {Acknowledgements :}}
The authors would like to thank  Rafael Hernandez, Kamal L. Panigrahi, Kentaroh Yoshida and Arkady Tseytlin  for their valuable comments on the manuscript. DR was supported through the Newton-Bhahba Fund and he would like to acknowledge the Royal Society UK and the Science and Engineering Research Board India (SERB) for financial assistance. AB acknowledges support  Fudan University and Thousand Young Talents Program. AB also acknowledges support form JSPS fellowship (P17023) and Yukawa Institute of Theoretical Physics. Aritra Banerjee (ArB) is supported in part by the Chinese Academy of Sciences (CAS) Hundred-Talent Program, by the Key Research Program of Frontier Sciences, CAS, and by Project 11647601 supported by NSFC. \\ 
%%%%%%%%%%%%%%%%%%%%%%%%%%%%%%%%%%%%%%%%%%%%%%%%%%%%%%%%%%%%%%%%%%%%%%%%%%%%%%%%%%%

\appendix
\section{SU(3) Coherent state:}
We give the details for our convention regarding the Gell -Mann matrices and the action  of $SU(3)$ coherent state that we have used in the main text. 
\begin{align}
\begin{split}
&\lambda_{1}= \left(
\begin{array}{ccc}
 0 & 1 & 0 \\
 1 & 0 & 0 \\
 0 & 0 & 0 \\
\end{array}
\right), \lambda_{2}= \left(
\begin{array}{ccc}
 0 & -i & 0 \\
 i & 0 & 0 \\
 0 & 0 & 0 \\
\end{array}
\right), \lambda_{3}= \left(
\begin{array}{ccc}
 1& 0 & 0 \\
 0 & -1 & 0 \\
 0 & 0 & 0 \\
\end{array}
\right),\\&\lambda_{4}= \left(
\begin{array}{ccc}
 0 & 0 & 1 \\
 0 & 0 & 0 \\
 1 & 0 & 0 \\
\end{array}
\right),\lambda_{5}= \left(
\begin{array}{ccc}
 0 & 0 & -i \\
 0 & 0 & 0 \\
 i & 0 & 0 \\
\end{array}
\right),\lambda_{6}=\left(
\begin{array}{ccc}
 0 & 0 & 0 \\
 0 & 0 & 1 \\
 0 & 1 & 0 \\
\end{array}
\right),\\&\lambda_{7}=\left(
\begin{array}{ccc}
 0 & 0 & 0 \\
 0 & 0 & -i \\
 0 & i & 0 \\
\end{array}
\right),\lambda_{8}=\left(
\begin{array}{ccc}
 \frac{1}{\sqrt{3}} & 0 & 0 \\
 0 & \frac{1}{\sqrt{3}} & 0 \\
 0 & 0 & -\frac{2}{\sqrt{3}} \\
\end{array}
\right).
\end{split}
\end{align}
 this we get,
\begin{align}
\begin{split}
&\langle n_{k}|\lambda_{1}|n_{k}\rangle=\cos (2 \varphi_k ) \cos ^2 \theta_k \sin (2 \psi_k ), ~\langle n_{k}|\lambda_{2}|n_{k}\rangle=\sin (2 \varphi_k ) \cos ^2 \theta_k \sin (2 \psi_k ),\\& \langle n_{k}|\lambda_{3}|n_{k}\rangle=\cos ^2 \theta_{k} \cos (2 \psi_k ),~ \langle n_{k}|\lambda_{4}|n_{k}\rangle=\sin (2 \theta_k ) \cos \psi_k \cos (\varphi_k -\phi_k ),\\& \langle n_{k}|\lambda_{5}|n_{k}\rangle=\sin (2 \theta_k ) \cos \psi_k \sin (\varphi_k -\phi_k ),~ \langle n_{k}|\lambda_{6}|n_{k}\rangle=\sin (2 \theta_k ) \sin \psi_k \cos (\varphi_k+\phi_k ),\\&\langle n_k|\lambda_{7}|n_{k}\rangle=-\sin (2 \theta_k ) \sin \psi_k \sin (\varphi_k+\phi_k ),~\langle n_k|\lambda_8|n_{k}\rangle= \frac{3 \cos (2 \theta_k )-1}{2 \sqrt{3}}.
\end{split}
\end{align}
Also it can be checked that given the coherent state (\ref{cohSU3}), using $\langle\vec n_{k}|\vec n_{k+1}\rangle$ we can reproduce the kinetic term in action (\ref{kappa3}) in the continuous limit.  
\section{The complex-$\b$ deformed theory}
In  \cite{Frolov:2005ty}, the gravity dual background of the complex $\beta$-deformed background corresponding to deformed $\mathcal{N}=4$ SYM theory takes the following form:
\begin{eqnarray}
\diff s^2&=& \sqrt{H} \left[ \diff s^2_{AdS_5}+\sum_{i=1}^3(\diff \rho_i^2+G \rho_i^2\diff \phi_i^2)
+(\g^2+\s^2) G \rho_1^2\rho_2^2\rho_3^2 \left(\sum_{i=1}^3 \diff \phi_i\right)^2 \right]\;,\\
B&=&(\g G w_2-12\s w_1 \diff\psi)\;, \quad \psi=\frac13(\phi_1+\phi_2+\phi_3),\\\no
w_2&=&\rho_1^2\rho_2^2\diff\phi_1\diff\phi_2-\rho_1^2\rho_3^2\diff\phi_1\diff\phi_3+\rho_2^2\rho_3^2\diff\phi_2\diff\phi_3, \quad \diff w_1=\cos\theta\sin^3\theta\sin\phi\cos\phi\diff\theta\diff\phi,\\\no
e^\Phi&=&e^{\Phi_0}\sqrt{G} H,
\end{eqnarray}
where the metric, the NS-NS $B$ field and the dilaton have been written. The functions used here are the following,
\begin{equation}
G=\frac{1}{1+(\g^2+\s^2) Q}\;,\quad \text{with}\quad  Q=\rho_1^2\rho_2^2+\rho_2^2\rho_3^2+\rho_1^2\rho_3^2,
\end{equation}
and
\begin{equation}
H=1+\s^2 Q\;,
\end{equation}
where the general complex $\beta$  can be written as,
\begin{equation}
\beta=\g-i \s;.
\end{equation}
The coordinates are related by the constraint  $\sum \rho_i^2=1$ and we choose to parametrise them in accordance with our $SU(3)$ coherent state
\begin{equation}
\rho_1=\cos\theta\cos\phi\;,\quad
\rho_2=\cos\theta\sin\phi\;,\quad
\rho_3=\sin\theta\;.
\end{equation}
The authors of  \cite{Puletti:2011hx} have considered a subsector of this theory with $\gamma= 0$, i.e. purely imaginary deformation parameter. The metric is indeed  same as that of real $\b$ but with the conformal factor $\sqrt{H}$ multiplied.

\section{A Bethe ansatz solution for energy in SU(3) case}
For the sake of completeness, let us now discuss the Bethe ansatz solution for the spectrum of circular spinning strings in the $SU(3)$ subsector. This is important since we will be using a $SU(3)_q$ Bethe ansatz which will hopefully be able to capture the deformed energy of the circular string in its entirety. In the usual $\mathcal{N} =4$ theory, the operators corresponding to these strings are supposed to have a generic form $Tr(\Phi_1^{J_1}\Phi_2^{J_2}\Phi_3^{J_3})$, where $J_i$ are the large R-charge and the long spin chain limit explicitly implies the large $J$ limit. Also as reported earlier, we consider $q=\text{exp}\Big[ -\frac{\kappa}{g\sqrt{1+\kappa^2}}\Big]=e^{-\nu/g}$. Following \cite{Puletti:2011hx}, one could write down the Bethe ansatz equations as,

\begin{equation}
\label{bethe1}
 \prod_{l\neq k}^{K_2} \frac{\sinh((\l_{2,k}-\l_{2,l})+\frac{\nu}{g})}{\sinh((\l_{2,k}-\l_{2,l})-\frac{\nu}{g})}\prod_{j=1}^{K_1} 
\frac{\sinh((\l_{2,k}-\l_{1,j})-\frac{\nu}{2g})}{\sinh ((\l_{2,k}-\l_{1,j})+\frac{\nu}{2g})}=1\,,
\end{equation}
\begin{equation}
\label{bethe2}
 \left(
\frac{\sinh (\l_{1,k}-\frac{\nu}{g})}{\sinh(\l_{1,k}+\frac{\nu}{g})}\right)^J=
\prod_{l\neq k}^{K_1} \frac{\sinh((\l_{1,k}-\l_{1,l})-\frac{\nu}{g})}{\sinh((\l_{1,k}-\l_{1,l})+\frac{\nu}{g})}\,
\prod_{j=1}^{K_2} \frac{\sinh((\l_{1,k}-\l_{2,j})+\frac{\nu}{2g})}{\sinh((\l_{1,k}-\l_{2,j})-\frac{\nu}{2g})}\,,
\quad 
\end{equation}
Here $J =J_1+J_2+J_3$ is the length of the spin chain. The periodicity condition imposes the constraint,
\begin{equation}
 \prod_{l}^{K_1} \frac{\sinh(\l_{1,l}-\frac{\nu}{g})}{\sinh(\l_{1,l}+\frac{\nu}{g})}=1\,,
\end{equation}
For exact values of $\nu$, the energy is given by
\begin{equation}
\label{bethe_energy1}
E=\sum_{k=1}^{K_1}\epsilon_k, \qquad \text{with} \qquad 
\epsilon_k=\frac{\lambda}{8\pi^2}
\frac{\sinh^2 \frac{\nu}{g}}{\sinh(\l_{1,k}+\frac{\nu}{2g})\sinh(\l_{1,k}-\frac{\nu}{2g})}\,.
\end{equation} 
For the fast spinning limit of the corresponding string solution in $J\to\infty$, one would have to take $\frac{\nu}{g}\to 0$ (which translates to $\kappa\to 0$), so that their product remains finite. The way to take this limit is take a logarithm of the Bethe equations and expand it accordingly. In this limit, the energy has an expression,
\begin{equation}
E= \frac{\lambda}{8\pi^2} \left(\sum_k^{K_1}\frac1{x_{1,k}^2}+\frac{\nu^2 J^2}{g^2}K_1 \right)\,,
\end{equation}
where $x_{m, k}$ is defined via $\tanh\lambda_{m,k} = 2iJx_{m,k}\tanh\frac{\nu}{2g}$, with $m=1,2$. After some strenuous algebra and identifying $J_1= J-K_1,~J_2= K_1-K_2,~J_3=K_2$ \cite{Freyhult:2005fn}, we can reach at the expression for energy which explicitly matches with (\ref{su3E}) when we identify $g = \frac{\l}{J^2}$.


\begin{thebibliography}{99}

\bibitem{Beisert1}
  N.~Beisert and P.~Koroteev,
 ``Quantum Deformations of the One-Dimensional Hubbard Model,''
  J.\ Phys.\ A {\bf 41} (2008) 255204
  doi:10.1088/1751-8113/41/25/255204
  [arXiv:0802.0777 [hep-th]].

\bibitem{Beisert2}
  N.~Beisert and F.~Spill,
 ``The Classical r-matrix of AdS/CFT and its Lie Bialgebra Structure,''
  Commun.\ Math.\ Phys.\  {\bf 285} (2009) 537
  doi:10.1007/s00220-008-0578-2
  [arXiv:0708.1762 [hep-th]].

\bibitem{Murgan}
  R.~Murgan and R.~I.~Nepomechie,
 ``q-deformed su(2$|$2) boundary S-matrices via the ZF algebra,''
  JHEP {\bf 0806} (2008) 096
  doi:10.1088/1126-6708/2008/06/096
  [arXiv:0805.3142 [hep-th]].
  
  
  \bibitem{Mansson1}
  T.~Mansson and K.~Zoubos,
 ``Quantum Symmetries and Marginal Deformations,''
  JHEP {\bf 1010} (2010) 043
  doi:10.1007/JHEP10(2010)043
  [arXiv:0811.3755 [hep-th]].

  \bibitem{Zoubos:2010kh}
  K.~Zoubos,
  ``Review of AdS/CFT Integrability, Chapter IV.2: Deformations, Orbifolds and Open Boundaries,''
  Lett.\ Math.\ Phys.\  {\bf 99} (2012) 375
  doi:10.1007/s11005-011-0515-8
  [arXiv:1012.3998 [hep-th]].

  
  \bibitem{vanTongeren:2013gva}
  S.~J.~van Tongeren,
  ``Integrability of the ${\rm Ad}{{{\rm S}}_{5}}\times {{{\rm S}}^{5}}$ superstring and its deformations,''
  J.\ Phys.\ A {\bf 47} (2014) 433001
  doi:10.1088/1751-8113/47/43/433001
  [arXiv:1310.4854 [hep-th]].
  
  \bibitem{Bena:2003wd}
  I.~Bena, J.~Polchinski and R.~Roiban,
  ``Hidden symmetries of the $AdS(5) x S**5$ superstring,''
  Phys.\ Rev.\ D {\bf 69} (2004) 046002
  doi:10.1103/PhysRevD.69.046002
  [hep-th/0305116].

\bibitem{Klimcik:2002zj}
  C.~Klimcik,
 ``Yang-Baxter sigma models and dS/AdS T duality,''
  JHEP {\bf 0212} (2002) 051
  doi:10.1088/1126-6708/2002/12/051
  [hep-th/0210095].
  
  \bibitem{Klimcik:2008eq}
  C.~Klimcik,
  ``On integrability of the Yang-Baxter sigma-model,''
  J.\ Math.\ Phys.\  {\bf 50} (2009) 043508
  doi:10.1063/1.3116242
  [arXiv:0802.3518 [hep-th]].
  
  \bibitem{Delduc:2013qra} 
  F.~Delduc, M.~Magro and B.~Vicedo,
  ``An integrable deformation of the $AdS_5 \times S^5$ superstring action,''
  Phys.\ Rev.\ Lett.\  {\bf 112}, no. 5, 051601 (2014)
  doi:10.1103/PhysRevLett.112.051601
  [arXiv:1309.5850 [hep-th]].
  
  \bibitem{Delduc:2014kha} 
  F.~Delduc, M.~Magro and B.~Vicedo,
  ``Derivation of the action and symmetries of the $q$-deformed $AdS_{5} \times S^{5}$ superstring,''
  JHEP {\bf 1410}, 132 (2014)
  doi:10.1007/JHEP10(2014)132
  [arXiv:1406.6286 [hep-th]].
  
\bibitem{Hoare:2014pna} 
  B.~Hoare, R.~Roiban and A.~A.~Tseytlin,
  ``On deformations of $AdS_n$ x $S^n$ supercosets,''
  JHEP {\bf 1406}, 002 (2014)
  doi:10.1007/JHEP06(2014)002
  [arXiv:1403.5517 [hep-th]].
 
 \bibitem{Arutyunov:2013ega} 
  G.~Arutyunov, R.~Borsato and S.~Frolov,
  ``S-matrix for strings on $\eta$-deformed AdS5 x S5,''
  JHEP {\bf 1404}, 002 (2014)
  doi:10.1007/JHEP04(2014)002
  [arXiv:1312.3542 [hep-th]].
  
  \bibitem{Arutynov:2014ota}
  G.~Arutyunov, M.~de Leeuw and S.~J.~van Tongeren,
  ``The exact spectrum and mirror duality of the $(\text{AdS}_5{\times}S^5)_\eta$ superstring,''
  Theor.\ Math.\ Phys.\  {\bf 182} (2015) no.1,  23
   [Teor.\ Mat.\ Fiz.\  {\bf 182} (2014) no.1,  28]
  doi:10.1007/s11232-015-0243-9
  [arXiv:1403.6104 [hep-th]].
  
  \bibitem{Arutyunov:2015qva} 
  G.~Arutyunov, R.~Borsato and S.~Frolov,
  ``Puzzles of $\eta$-deformed AdS$_5 \times$ S$^5$,''
  JHEP {\bf 1512}, 049 (2015)
  doi:10.1007/JHEP12(2015)049
  [arXiv:1507.04239 [hep-th]].
  
  \bibitem{Engelund:2014pla} 
  O.~T.~Engelund and R.~Roiban,
  ``On the asymptotic states and the quantum S matrix of the $\eta$-deformed AdS$_{5} \times$ S$^{5}$ superstring,''
  JHEP {\bf 1503}, 168 (2015)
  doi:10.1007/JHEP03(2015)168
  [arXiv:1412.5256 [hep-th]].
  
  \bibitem{Lunin:2014tsa} 
  O.~Lunin, R.~Roiban and A.~A.~Tseytlin,
  ``Supergravity backgrounds for deformations of AdS$_{n} \times S^n$ supercoset string models,''
  Nucl.\ Phys.\ B {\bf 891}, 106 (2015)
  doi:10.1016/j.nuclphysb.2014.12.006
  [arXiv:1411.1066 [hep-th]].
  

\bibitem{Kawaguchi:2014qwa} 
  I.~Kawaguchi, T.~Matsumoto and K.~Yoshida,
  ``Jordanian deformations of the $AdS_5 x S^5$ superstring,''
  JHEP {\bf 1404}, 153 (2014)
  doi:10.1007/JHEP04(2014)153
  [arXiv:1401.4855 [hep-th]].


\bibitem{Kawaguchi:2014fca} 
  I.~Kawaguchi, T.~Matsumoto and K.~Yoshida,
  ``A Jordanian deformation of AdS space in type IIB supergravity,''
  JHEP {\bf 1406}, 146 (2014)
  doi:10.1007/JHEP06(2014)146
  [arXiv:1402.6147 [hep-th]].

\bibitem{gandu}
  T.~Araujo, E.~O Colgain, J.~Sakamoto, M.~M.~Sheikh-Jabbari and K.~Yoshida,
 ``$I$ in generalized supergravity,''
  Eur.\ Phys.\ J.\ C {\bf 77} (2017) no.11,  739
  doi:10.1140/epjc/s10052-017-5316-5
  [arXiv:1708.03163 [hep-th]].

\bibitem{gandu1}
  I.~Bakhmatov, O.~Kelekci, E.~O Colgain and M.~M.~Sheikh-Jabbari,
 ``Classical Yang-Baxter Equation from Supergravity,''
  arXiv:1710.06784 [hep-th].

  
  \bibitem{Blau:2001ne} 
  M.~Blau, J.~M.~Figueroa-O'Farrill, C.~Hull and G.~Papadopoulos,
  ``A New maximally supersymmetric background of IIB superstring theory,''
  JHEP {\bf 0201}, 047 (2002)
  doi:10.1088/1126-6708/2002/01/047
  [hep-th/0110242].
  
  \bibitem{Metsaev:2001bj} 
  R.~R.~Metsaev,
  ``Type IIB Green-Schwarz superstring in plane wave Ramond-Ramond background,''
  Nucl.\ Phys.\ B {\bf 625}, 70 (2002)
  doi:10.1016/S0550-3213(02)00003-2
  [hep-th/0112044].
  
  \bibitem{Metsaev:2002re} 
  R.~R.~Metsaev and A.~A.~Tseytlin,
  ``Exactly solvable model of superstring in Ramond-Ramond plane wave background,''
  Phys.\ Rev.\ D {\bf 65}, 126004 (2002)
  doi:10.1103/PhysRevD.65.126004
  [hep-th/0202109].
  
  \bibitem{Berenstein:2002jq} 
  D.~E.~Berenstein, J.~M.~Maldacena and H.~S.~Nastase,
  ``Strings in flat space and pp waves  N=4 superYang-Mills,''
  JHEP {\bf 0204}, 013 (2002)
  doi:10.1088/1126-6708/2002/04/013
  [hep-th/0202021].
  
  \bibitem{Callan:2003xr} 
  C.~G.~Callan, Jr., H.~K.~Lee, T.~McLoughlin, J.~H.~Schwarz, I.~Swanson and X.~Wu,
  ``Quantizing string theory in AdS(5) x S**5: Beyond the pp wave,''
  Nucl.\ Phys.\ B {\bf 673}, 3 (2003)
  doi:10.1016/j.nuclphysb.2003.09.008
  [hep-th/0307032].
  
  \bibitem{Callan:2004uv} 
  C.~G.~Callan, Jr., T.~McLoughlin and I.~Swanson,
  ``Holography beyond the Penrose limit,''
  Nucl.\ Phys.\ B {\bf 694}, 115 (2004)
  doi:10.1016/j.nuclphysb.2004.06.033
  [hep-th/0404007].
  
  \bibitem{Mizoguchi:2002qy} 
  S.~Mizoguchi, T.~Mogami and Y.~Satoh,
  ``Penrose limits and Green-Schwarz strings,''
  Class.\ Quant.\ Grav.\  {\bf 20}, 1489 (2003)
  doi:10.1088/0264-9381/20/8/306
  [hep-th/0209043].
  
 \bibitem{PandoZayas:2002dso} 
  L.~A.~Pando Zayas and J.~Sonnenschein,
  ``On Penrose limits and gauge theories,''
  JHEP {\bf 0205}, 010 (2002)
  doi:10.1088/1126-6708/2002/05/010
  [hep-th/0202186].
  
  \bibitem{Plefka:2003nb} 
  J.~C.~Plefka,
  ``Lectures on the plane wave string / gauge theory duality,''
  Fortsch.\ Phys.\  {\bf 52}, 264 (2004)
  doi:10.1002/prop.200310121
  [hep-th/0307101].
  
  \bibitem{Beisert:2003ea} 
  N.~Beisert, S.~Frolov, M.~Staudacher and A.~A.~Tseytlin,
  ``Precision spectroscopy of AdS / CFT,''
  JHEP {\bf 0310}, 037 (2003)
  doi:10.1088/1126-6708/2003/10/037
  [hep-th/0308117].
  

\bibitem{Arutyunov:2003uj}
  G.~Arutyunov, S.~Frolov, J.~Russo and A.~A.~Tseytlin,
  ``Spinning strings in  $AdS(5) x S**5$ and integrable systems,''
  Nucl.\ Phys.\ B {\bf 671} (2003) 3
  doi:10.1016/j.nuclphysb.2003.08.036
  [hep-th/0307191].
 
  
  \bibitem{Kruczenski:2003gt} 
  M.~Kruczenski,
  ``Spin chains and string theory,''
  Phys.\ Rev.\ Lett.\  {\bf 93}, 161602 (2004)
  doi:10.1103/PhysRevLett.93.161602
  [hep-th/0311203].
  
  \bibitem{Arutyunov:2003za}
  G.~Arutyunov, J.~Russo and A.~A.~Tseytlin,
  ``Spinning strings in $AdS(5) x S**5$: New integrable system relations,''
  Phys.\ Rev.\ D {\bf 69} (2004) 086009
  doi:10.1103/PhysRevD.69.086009
  [hep-th/0311004].

  
  \bibitem{Kruczenski:2004kw} 
  M.~Kruczenski, A.~V.~Ryzhov and A.~A.~Tseytlin,
  ``Large spin limit of AdS(5) x S**5 string theory and low-energy expansion of ferromagnetic spin chains,''
  Nucl.\ Phys.\ B {\bf 692}, 3 (2004)
  doi:10.1016/j.nuclphysb.2004.05.028
  [hep-th/0403120].
  
  

  
  \bibitem{Hernandez:2004uw} 
  R.~Hernandez and E.~Lopez,
  ``The SU(3) spin chain sigma model and string theory,''
  JHEP {\bf 0404}, 052 (2004)
  doi:10.1088/1126-6708/2004/04/052
  [hep-th/0403139].
  
  \bibitem{Stefanski:2004cw} 
  B.~Stefanski, Jr. and A.~A.~Tseytlin,
  ``Large spin limits of AdS/CFT and generalized Landau-Lifshitz equations,''
  JHEP {\bf 0405}, 042 (2004)
  doi:10.1088/1126-6708/2004/05/042
  [hep-th/0404133].
  
  \bibitem{Minahan:2005mx} 
  J.~A.~Minahan, A.~Tirziu and A.~A.~Tseytlin,
  ``1/J corrections to semiclassical AdS/CFT states  quantum Landau-Lifshitz model,''
  Nucl.\ Phys.\ B {\bf 735}, 127 (2006)
  doi:10.1016/j.nuclphysb.2005.12.003
  [hep-th/0509071].
  
  \bibitem{Minahan:2005qj} 
  J.~A.~Minahan, A.~Tirziu and A.~A.~Tseytlin,
  ``1/J**2 corrections to BMN energies  the quantum long range Landau-Lifshitz model,''
  JHEP {\bf 0511}, 031 (2005)
  doi:10.1088/1126-6708/2005/11/031
  [hep-th/0510080].
  
  \bibitem{Kazakov:2004qf} 
  V.~A.~Kazakov, A.~Marshakov, J.~A.~Minahan and K.~Zarembo,
  ``Classical/quantum integrability in AdS/CFT,''
  JHEP {\bf 0405}, 024 (2004)
  doi:10.1088/1126-6708/2004/05/024
  [hep-th/0402207].
  
  \bibitem{Frolov:2005ty} 
  S.~A.~Frolov, R.~Roiban and A.~A.~Tseytlin,
  ``Gauge-string duality for superconformal deformations of N=4 super Yang-Mills theory,''
  JHEP {\bf 0507}, 045 (2005)
  doi:10.1088/1126-6708/2005/07/045
  [hep-th/0503192].
  
  \bibitem{Wen:2006fw} 
  W.~Y.~Wen,
  ``Spin chain  marginally deformed AdS(3) x S**3,''
  Phys.\ Rev.\ D {\bf 75}, 067901 (2007)
  doi:10.1103/PhysRevD.75.067901
  [hep-th/0610147].
  
  \bibitem{Dimov:2004qv} 
  H.~Dimov and R.~C.~Rashkov,
  ``A Note on spin chain / string duality,''
  Int.\ J.\ Mod.\ Phys.\ A {\bf 20}, 4337 (2005)
  doi:10.1142/S0217751X05020975
  [hep-th/0403121].
  
  \bibitem{Bellucci:2004qr} 
  S.~Bellucci, P.-Y.~Casteill, J.~F.~Morales and C.~Sochichiu,
  ``SL(2) spin chain and spinning strings on AdS(5) x S**5,''
  Nucl.\ Phys.\ B {\bf 707}, 303 (2005)
  doi:10.1016/j.nuclphysb.2004.11.020
  [hep-th/0409086].
  
  \bibitem{Huang:2006bh} 
  W.~H.~Huang,
  ``Spin Chain with Magnetic Field and Spinning String in Magnetic Field Background,''
  Phys.\ Rev.\ D {\bf 74}, 027901 (2006)
  doi:10.1103/PhysRevD.74.027901
  [hep-th/0605242].
  
  \bibitem{Dorey:2008zy} 
  N.~Dorey,
  ``A Spin Chain  String Theory,''
  Acta Phys.\ Polon.\ B {\bf 39}, 3081 (2008)
  [arXiv:0805.4387 [hep-th]].
  
  \bibitem{Kameyama:2014bua} 
  T.~Kameyama and K.~Yoshida,
  ``Anisotropic Landau-Lifshitz sigma models  $q$-deformed AdS$_{5} \times$ S$^{5}$ superstrings,''
  JHEP {\bf 1408}, 110 (2014)
  doi:10.1007/JHEP08(2014)110
  [arXiv:1405.4467 [hep-th]].

\bibitem{Arutyunov:2014cda}
  G.~Arutyunov and D.~Medina-Rincon,
  ``Deformed Neumann model from spinning strings on ($AdS_5 \times S^5$)$_\eta$,''
  JHEP {\bf 1410} (2014) 050
  doi:10.1007/JHEP10(2014)050
  [arXiv:1406.2536 [hep-th]].
  
  \bibitem{Arutyunov:2016ysi}
  G.~Arutyunov, M.~Heinze and D.~Medina-Rincon,
  ``Integrability of the $\eta$-deformed Neumann?Rosochatius model,''
  J.\ Phys.\ A {\bf 50} (2017) no.3,  035401
  doi:10.1088/1751-8121/50/3/035401
  [arXiv:1607.05190 [hep-th]].

\bibitem{Kameyama:2014vma}
  T.~Kameyama and K.~Yoshida,
  ``A new coordinate system for $q$-deformed AdS$_{5} \times$ S$^5$ and classical string solutions,''
  J.\ Phys.\ A {\bf 48} (2015) no.7,  075401
  doi:10.1088/1751-8113/48/7/075401
  [arXiv:1408.2189 [hep-th]].
  
  \bibitem{Hernandez1}
  R.~Hernandez and J.~M.~Nieto,
  ``Spinning strings in the $\eta$-deformed Neumann-Rosochatius system,''
  Phys.\ Rev.\ D {\bf 96} (2017) no.8,  086010
  doi:10.1103/PhysRevD.96.086010
  [arXiv:1707.08032 [hep-th]].

  \bibitem{Kameyama:2013qka} 
  T.~Kameyama and K.~Yoshida,
  ``String theories on warped AdS backgrounds and integrable deformations of spin chains,''
  JHEP {\bf 1305}, 146 (2013)
  doi:10.1007/JHEP05(2013)146
  [arXiv:1304.1286 [hep-th]].
  
  \bibitem{Frolov:2005iq} 
  S.~A.~Frolov, R.~Roiban and A.~A.~Tseytlin,
  ``Gauge-string duality for (non)supersymmetric deformations of N=4 super Yang-Mills theory,''
  Nucl.\ Phys.\ B {\bf 731}, 1 (2005)
  doi:10.1016/j.nuclphysb.2005.10.004
  [hep-th/0507021].
  
  \bibitem{Lunin:2005jy}
  O.~Lunin and J.~M.~Maldacena,
  ``Deforming field theories with $U(1) \times U(1)$ global symmetry and their gravity duals,''
  JHEP {\bf 0505} (2005) 033
  doi:10.1088/1126-6708/2005/05/033
  [hep-th/0502086].


\bibitem{Frolov:2005dj}
  S.~Frolov,
  ``Lax pair for strings in Lunin-Maldacena background,''
  JHEP {\bf 0505} (2005) 069
  doi:10.1088/1126-6708/2005/05/069
  [hep-th/0503201].

  \bibitem{Perelomov:1986tf} 
  A.~M.~Perelomov,
  ``Generalized coherent states and their applications,''
  Berlin, Germany: Springer (1986) 320 p

  
  \bibitem{Dimov:2004qv} 
  H.~Dimov and R.~C.~Rashkov,
  ``A Note on spin chain / string duality,''
  Int.\ J.\ Mod.\ Phys.\ A {\bf 20}, 4337 (2005)
  doi:10.1142/S0217751X05020975
  [hep-th/0403121].
    
  
\bibitem{Ryang:2004pu} 
  S.~Ryang,
  ``Circular and folded multi-spin strings in spin chain sigma models,''
  JHEP {\bf 0410}, 059 (2004)
  doi:10.1088/1126-6708/2004/10/059
  [hep-th/0409217].

  

\bibitem{Lamers:2015dfa}
  J.~Lamers,
 ``A pedagogical introduction to quantum integrability, with a view towards theoretical high-energy physics,''
  PoS Modave {\bf 2014} (2015) 001
  [arXiv:1501.06805 [math-ph]].

\bibitem{Kristjansen:2004ei}
  C.~Kristjansen,
  ``Three spin strings on $AdS(5) x S**5$  N=4 SYM,''
  Phys.\ Lett.\ B {\bf 586} (2004) 106
  doi:10.1016/j.physletb.2004.02.037
  [hep-th/0402033].
  
 
  
  \bibitem{Kristjansen:2004za}
  C.~Kristjansen and T.~Mansson,
 ``The Circular, elliptic three spin string  the SU(3) spin chain,''
  Phys.\ Lett.\ B {\bf 596} (2004) 265
  doi:10.1016/j.physletb.2004.06.099
  [hep-th/0406176].

\bibitem{Freyhult:2005fn}
  L.~Freyhult and C.~Kristjansen,
  ``Rational three-spin string duals and non-anomalous finite size effects,''
  JHEP {\bf 0505} (2005) 043
  doi:10.1088/1126-6708/2005/05/043
  [hep-th/0502122].

\bibitem{Fateev:1996ea}
  V.~A.~Fateev,
  ``The sigma model (dual) representation for a two-parameter family of integrable quantum field theories,''
  Nucl.\ Phys.\ B {\bf 473} (1996) 509.
  doi:10.1016/0550-3213(96)00256-8
  
  
    \bibitem{Leigh:1995ep}
  R.~G.~Leigh and M.~J.~Strassler,
  ``Exactly marginal operators and duality in four-dimensional N=1 supersymmetric gauge theory,''
  Nucl.\ Phys.\ B {\bf 447} (1995) 95
  doi:10.1016/0550-3213(95)00261-P
  [hep-th/9503121].
  
  
\bibitem{Bundzik:2005zg} 
  D.~Bundzik and T.~Mansson,
  ``The General Leigh-Strassler deformation and integrability,''
  JHEP {\bf 0601}, 116 (2006)
  doi:10.1088/1126-6708/2006/01/116
  [hep-th/0512093].
  

\bibitem{Matsumoto:2014nra} 
  T.~Matsumoto and K.~Yoshida,
  ``Lunin-Maldacena backgrounds  the classical Yang-Baxter equation - towards the gravity/CYBE correspondence,''
  JHEP {\bf 1406}, 135 (2014)
  doi:10.1007/JHEP06(2014)135
  [arXiv:1404.1838 [hep-th]].
  
  
\bibitem{Puletti:2011hx}
  V.~G.~M.~Puletti and T.~Mansson,
  ``The dual string sigma-model of the $SU_q(3)$ sector,''
  JHEP {\bf 1201} (2012) 129
  doi:10.1007/JHEP01(2012)129
  [arXiv:1106.1116 [hep-th]].

\bibitem{Mansson:2007sh} 
  T.~Mansson,
  ``The Leigh-Strassler Deformation and the Quest for Integrability,''
  JHEP {\bf 0706}, 010 (2007)
  doi:10.1088/1126-6708/2007/06/010
  [hep-th/0703150].
 


  
  
\bibitem{Giataganas:2013dha}
  D.~Giataganas, L.~A.~Pando Zayas and K.~Zoubos,
  ``On Marginal Deformations and Non-Integrability,''
  JHEP {\bf 1401} (2014) 129
  doi:10.1007/JHEP01(2014)129
  [arXiv:1311.3241 [hep-th]].
    
\bibitem{Roychowdhury:2017vdo}
  D.~Roychowdhury,
  ``Analytic integrability for strings on $ \eta $ and $ \lambda $ deformed backgrounds,''
  JHEP {\bf 1710} (2017) 056
  doi:10.1007/JHEP10(2017)056
  [arXiv:1707.07172 [hep-th]].
    
  \bibitem{Hernandez:2004kr}
  R.~Hernandez and E.~Lopez,
  ``Spin chain sigma models with fermions,''
  JHEP {\bf 0411} (2004) 079
  doi:10.1088/1126-6708/2004/11/079
  [hep-th/0410022].
  
  \bibitem{Stefanski:2005tr}
  B.~Stefanski, Jr. and A.~A.~Tseytlin,
  ``Super spin chain coherent state actions and AdS(5) x S**5 superstring,''
  Nucl.\ Phys.\ B {\bf 718} (2005) 83
  doi:10.1016/j.nuclphysb.2005.04.026
  [hep-th/0503185].
  
  \bibitem{Stefanski:2007dp}
  B.~Stefanski, Jr.,
  ``Landau-Lifshitz sigma-models, fermions and the AdS/CFT correspondence,''
  JHEP {\bf 0707} (2007) 009
  doi:10.1088/1126-6708/2007/07/009
  [arXiv:0704.1460 [hep-th]].
  
  \bibitem{Arnaudon:1997ej} 
  D.~Arnaudon,
  ``Algebraic approach to q deformed supersymmetric variants of the Hubbard model with pair hoppings,''
  JHEP {\bf 9712}, 006 (1997)
  doi:10.1088/1126-6708/1997/12/006
  
  \bibitem{Bazhanov:2008yc} 
  V.~V.~Bazhanov and Z.~Tsuboi,
  ``Baxter's Q-operators for supersymmetric spin chains,''
  Nucl.\ Phys.\ B {\bf 805}, 451 (2008)
  doi:10.1016/j.nuclphysb.2008.06.025
  [arXiv:0805.4274 [hep-th]].
  
 \end{thebibliography}
\end{document}